\documentclass[prd, 10pt,aps,nofootinbib, floatfix, notitlepage,twocolumn,longbibliography]{revtex4-2}

\usepackage{amsmath,amsfonts,amssymb,fontawesome5}
\usepackage[normalem]{ulem}
\usepackage{mathrsfs}
\usepackage{graphicx}
\usepackage[english]{babel} 
\usepackage{url} 
\usepackage{color}
\usepackage{xcolor}
\usepackage{bbold}
\usepackage{adjustbox}
\usepackage[utf8]{inputenc} 
\usepackage[colorlinks=true,citecolor=blue,urlcolor=blue]{hyperref}
\newcommand{\be}{\begin{equation}}
\newcommand{\ee}{\end{equation}}
\newcommand{\bes}{\begin{equation*}}
\newcommand{\ees}{\end{equation*}}

\renewcommand{\vec}[1]{\boldsymbol{#1}}

\begin{document}

\title{
The Phonon Background from Gamma Rays in Sub-GeV Dark Matter Detectors
}
\author{Kim V. Berghaus$^\dagger$}
\author{Rouven Essig$^\dagger$}
\author{Yonit Hochberg$^\star$}
\author{Yutaro Shoji$^\star$}
\author{Mukul Sholapurkar$^{\dagger,\ddagger}$}

\affiliation{$^\dagger$C.N. Yang Institute for Theoretical Physics, Stony Brook University, NY 11794, USA}
\affiliation{$^\star$Racah Institute of Physics, Hebrew University of Jerusalem, Jerusalem 91904, Israel} 
\affiliation{$^\ddagger$Department of Physics, University of California, San Diego, CA 92093, USA}

\begin{abstract}
 High-energy photons with $\mathcal{O}$(MeV) energies from radioactive contaminants can scatter in a solid-state target material and constitute an important low-energy background for sub-GeV dark matter direct-detection searches.  This background is most noticeable for energy deposits in the $1 - 100$~meV range due to the partially coherent scattering enhancement in the forward scattering direction. We comprehensively quantify the resulting single- and multi-phonon background in Si, Ge, GaAs, SiC, and Al$_2$O$_3$ target materials, which are representative of target materials of interest in low-mass dark matter searches.  We use a realistic representation of the high-energy photon background, and contrast the expected background phonon spectrum with the expected dark matter signal phonon spectrum.  An active veto is needed to suppress this background sufficiently in order to allow for the detection of a dark matter signal, even in well-shielded environments. For comparison we also show the expected single- and multi-phonon event rates from coherent neutrino-nucleus scattering due to solar neutrinos, and find that they are sub-dominant to the photon-induced phonon background.  \href{https://github.com/KBerghaus/phonon_background}{\faGithub} 

\end{abstract}

\date{\today}

\maketitle

\section{Introduction}
\label{sec:intro}

The nature of dark matter (DM) is one of the biggest open research questions in fundamental physics. Growing efforts to directly detect non-gravitational particle-like interactions of DM with masses lighter than the proton have inspired a multitude of upcoming and expanding experiments over the last decade~\cite{Angle:2011th,Essig:2012yx,Hochberg:2015pha,Hochberg:2015fth,Aprile:2016wwo,Schutz:2016tid,Essig:2017kqs,EDELWEISS:2017lvq,Tiffenberg:2017aac,Angloher:2017sxg,Romani:2017iwi,Hochberg:2017wce,Cavoto:2017otc,Agnes:2018oej,CDMS:2018nhx,Agnese:2018col,Aguilar-Arevalo:2019wdi,Essig:2019kfe,Physicality.125.171802,Aprile:2019xxb,Amaral:2020ryn,XENON:2019gfn,Hochberg:2019cyy,Abdelhameed:2019hmk,Abdelhameed:2019mac,Blanco:2019lrf,SENSEI:2020dpa,LUX:2020vbj,Arnaud:2020svb,Canonica:2020omq,SuperCDMS:2020aus,Hochberg:2021yud,Chiles:2021gxk,SPICE}. 
One promising such direction is the use of semiconductor detectors, which can be sensitive to light DM in the $10\, \text{keV}- \text{100 \,\text{MeV}}$ mass range through electron and phonon signals (see, {\it e.g.},~\cite{Essig:2011nj,Graham:2012su,Essig:2015cda,Lee:2015qva,Hochberg:2016ntt,Derenzo:2016fse,Knapen:2017ekk,Griffin:2018bjn,Kurinsky:2019pgb, Griffin:2019mvc,Caputo:2019cyg,Trickle:2019nya, Trickle:2020oki,Griffin:2020lgd,Griffin:2021znd,Coskuner:2021qxo}). Characterizing, quantifying, and mitigating the backgrounds in these experiments is a crucial task to ensure their success~\cite{Robinson_2017,Du:2020ldo,Robinson:2020zec}.

High-energy ($\sim 100~\text{keV}-\text{few MeV}$) photons are produced in DM detectors from the decay of radioactive impurities in the detector or shielding material. These photons have a large scattering cross-section at small momentum transfers~\cite{Robinson_2017} and can scatter in the target material to produce a single- or multi-phonon excitation, mimicking a low-mass DM signal. Such low-energy ($\sim 1-100~\text{meV}$) excitations could be within the reach of upcoming technologies such as transition edge sensors and form the basis of R\&D efforts for detecting sub-GeV DM.  It is therefore imperative that these backgrounds are characterized carefully.  In this work, we derive the single- and multi-phonon signals from the scattering of high-energy photons in several semiconductor targets (Si, Ge, GaAs, SiC, and Al$_2$O$_3$).\footnote{Phonons can also be created by absorbing infrared photons, which could be generated from Cherenkov radiation or from the radiative recombination of electron-hole pairs created by high-energy charged particles~\cite{Du:2020ldo}.  In this work, however, we focus on the phonon signal from high-energy photons.} We contrast the expected background spectrum with the expected signal spectrum from several DM candidates.  We further estimate the photon-veto efficiencies with the detector, and discuss the consequences of this background on the sensitivity reach of proposed low-mass DM searches. 

The outline of this paper is as follows. In Sec.~\ref{sec:analysis} we review all processes giving rise to photon-ion scattering, and identify the dominant one as photon-electron Rayleigh scattering. In Sec.~\ref{sec:analysis2} we show how to use a material's phonon density of states to calculate the dynamic structure function for single- and multi-phonon events. In Sec.~\ref{sec:results} we present the total background spectrum between $\omega \sim 1 \, \text{meV} - 100\, \text{meV}$ for a variety of materials, and contrast our expected background with three different DM signal models in GaAs. Finally, in Sec.~\ref{sec:disc} we briefly summarize our main findings and conclude that an active veto is needed to sufficiently suppress the photon background.  Several Appendices provide additional details. 

\section{Photon-Ion Scattering}
\label{sec:analysis}

Much of the theoretical framework of photon-atom scattering is applicable to photon scattering in condensed matter materials; however, due to the non-local nature of valence electrons in semiconductors this framework only extends to the tightly bound core electrons. The outer-shell electrons, however, do not contribute much to photon-atom scattering at the momentum transfers of interest ($\gtrsim 10-100$~keV), so we can reliably approximate photon-atom scattering as photon-ion scattering.  Thus, in the following discussion, we consider photon-ion scattering and only include the charges of the tightly bound electrons.

We begin by deriving the differential background rate, $\frac{dR}{d\omega}(\omega)$, as a function of energy deposition $\omega$, for $\omega\sim 1 - 100$~meV for high-energy photons scattering off a target material consisting of Si, Ge, GaAs, SiC, or $\text{Al}_2\text{O}_3$. We calculate the rate as a function of the background photon number densities $n_{\gamma_i}$, with discrete photon energies $E_{\gamma_i}$, where the subscript $i$ denotes a particular photon energy, and the double differential photon-ion scattering cross-section, $\frac{d\sigma}{d\Omega d\omega}$, 
\begin{equation} \label{rate}
\frac{dR}{d\omega}(\omega)  = N_T \sum_{i} \int d\Omega  \frac{d\sigma}{d\Omega d\omega}({\bf q}, E_{\gamma_i},\omega) n_{\gamma_i} \,.
\end{equation}
Here ${\bf q}$ is the momentum transfer of the photon to the ion and $N_T$ the number of target atoms. The double-differential cross-section $\frac{d\sigma}{d\Omega d\omega}$ can be factorized into the differential 
partially coherent photon-single-ion scattering cross section times a dynamic structure function $S({\bf q},\omega)$ (sometimes called the partial dynamic structure function), which captures the target-specific material response to a given energy-momentum deposition, 
\begin{equation} \label{xsection}
\frac{d\sigma}{d\Omega d\omega}({\bf q}, E_{\gamma},\omega) = \frac{d\sigma}{d\Omega}({\bf q}, E_{\gamma}) S({\bf q}, \omega) \,.
\end{equation}
The dynamic structure function $S({\bf q},\omega)$ is determined by the accessible degrees of freedom in the material at energy depositions of $\omega$. For energies $\omega$ below $\sim1 \, \text{eV}$, the relevant material excitation channels are vibrational degrees of freedom, namely single-phonon and multi-phonon states. We refer to momentum transfers up to $q \sim {\cal O}(100~\text{keV})$ as partially coherent scatterings, where the partial coherence refers to coherence over an individual ion ($r_{\text{ion}}^{-1} \sim {\cal O}(1~{\rm keV)}$).
We discuss here the photon-ion scattering, and leave the discussion of the dynamic structure function and the generation of phonons to Sec.~\ref{sec:analysis2}. 

The partially coherent photon-single-ion differential cross section,  $d\sigma/d\Omega$, can be expressed in terms of the parallel ($\parallel$) and perpendicular ($\perp$) polarization states ($\lambda$) of the incoming photon as~\cite{KANE198675, doi:10.1063/1.556027, ROY19993} 
\begin{equation}
    \frac{d\sigma}{d\Omega}=\frac{\alpha^2}{2m_e^2}\sum_{\lambda=\perp,\parallel}|A_\lambda(E_\gamma,\theta)|^2,
\end{equation}
where $E_\gamma$ is the incident photon energy, $\theta$ is the scattering angle, and 
\begin{align}
    A_\lambda(E_\gamma,\theta)&=A_\lambda^{\rm R}+A_\lambda^{\rm N}+A_\lambda^{\rm D}+A_\lambda^{\rm NR}.
\end{align}
The first term, $A_\lambda^{\rm R}$, is the contribution of the photon-electron Rayleigh scattering; the second term, $A^N_{\lambda}$, is that of nuclear Thomson scattering; $A^{\rm D}_\lambda$ denotes
Delbr\"uck scattering; and the last term, $A^{\rm NR}_\lambda$, is that of the nuclear resonance scattering. We now discuss each of these terms in turn.  As we will see, the electron Rayleigh scattering dominates for the energies and momentum-transfers of interest. 

\begin{figure*}[!t]
    \adjustimage{width=0.9\textwidth,center}{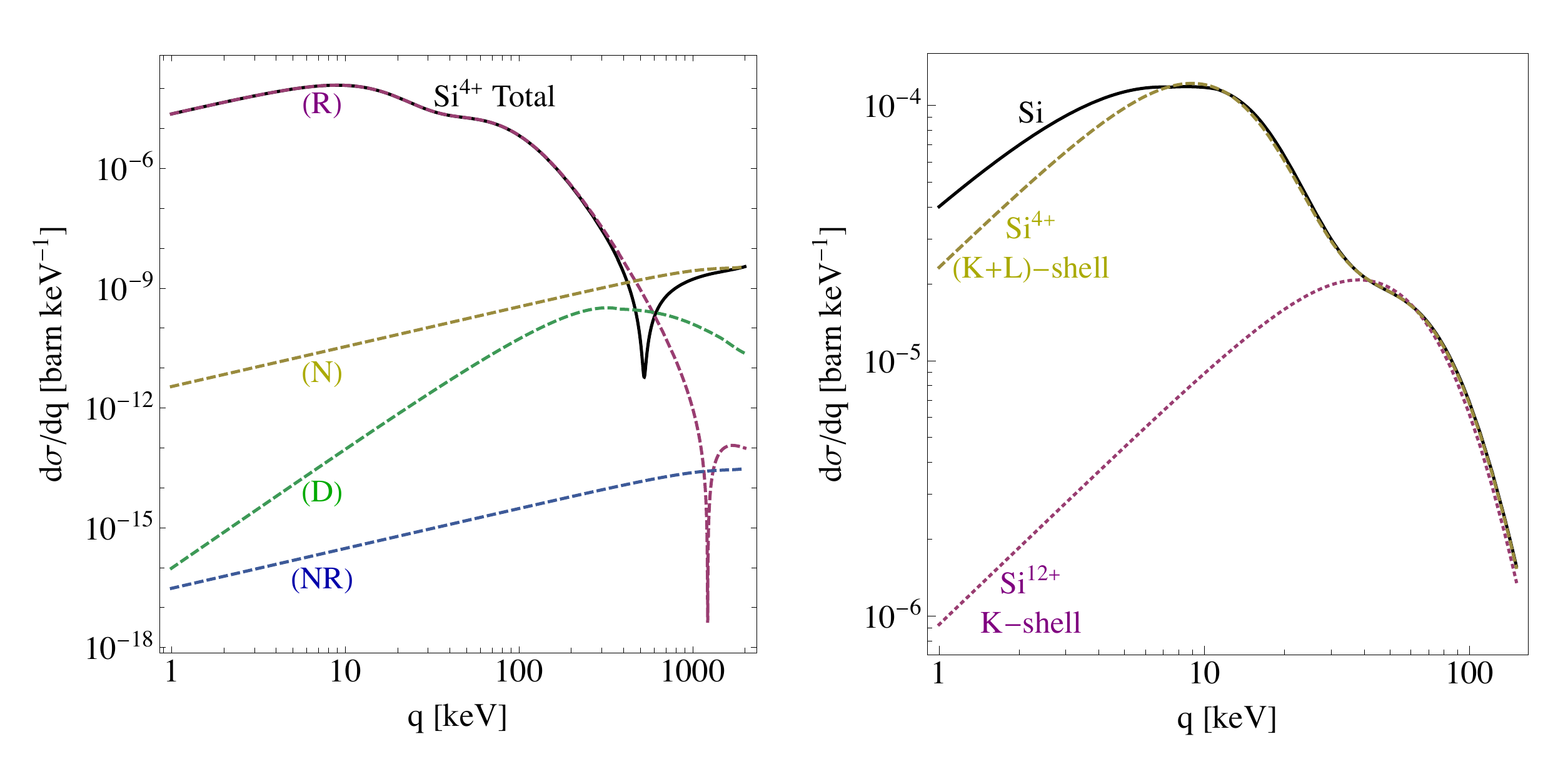}
    \vspace{-0.7cm}
    \caption{Total photon-ion scattering cross section and its respective contributions for a Si ion, for a photon energy of $E_\gamma=1461~{\rm keV}$. In the left panel, the black solid line shows the total cross section. The purple, yellow, green and blue dashed lines indicate the photon-electron Rayleigh (R), the nuclear Thomson (N), the Delbr\"uck (D) and the nuclear resonance (NR) contributions, respectively. In the right panel, we show the photon-electron Rayleigh scattering cross section,  with the innermost (K-shell) electrons (purple dotted), with the tightly-bound (K- and L-shell)  electrons (yellow dashed), and with all electrons (black solid). The yellow-dashed line corresponds to photon-Si-ion scattering, while the black line corresponds to photon-Si-atom scattering; we see that the cross-sections agree for $q\gtrsim 10$~keV.} 
     \label{fig:contr}
\end{figure*}

The photon-electron Rayleigh scattering amplitude is given by
\begin{equation}
    A_\perp^{\rm R}=g(q)\,,\quad ~A_\parallel^{\rm R}=A_\perp^{\rm R}\cos\theta\,,
\end{equation}
where
\begin{equation} \label{qtheta}
    q=2E_\gamma\sin\frac{\theta}{2}\,, 
\end{equation}
and $g(q)$ is the form factor of the photon-electron Rayleigh scattering.  For $g(q)$, we adopt the so-called modified form factor,\footnote{One may add a $\theta$-independent anomalous form factor term to $g(q)$ (see {\it e.g.}~\cite{Kissel:st0621}), but this term only becomes important at much higher momentum transfers than of interest.}

\begin{equation} \label{mformfactor}
    g(q)=\sum_{j=1}^{Z_{\rm ion}}\int_0^\infty dr\frac{\sin qr}{qr}\rho_j(q)\frac{m_e}{m_e+E_j-V(r)}\,.
\end{equation}

This expression is comprised of the standard atomic form factor (the Fourier transform of the electron charge density distribution)  and a relativistic correction factor $\frac{m_e}{m_e+E_j-V(r)}$. Here, $Z_{\rm ion}$ is the number of electrons in an ion, $\rho_j(r)$ and $E_j<0$ are the charge density and the energy eigenvalue of the $j$-th electron, and $V(r)$ is the Hartree-Fock potential for the electron. For the recoil energies of our interest, $\omega\sim 1~{\rm meV}-100~{\rm meV}$, where phonon and multi-phonons arise in condensed matter systems, the relevant momentum transfer is $q \sim 10~{\rm keV}-100~{\rm keV}$ for a silicon atom. For this energy range, the scattering is partially coherent over the ion (it would be fully coherent for $q\ll \,$1~keV).  We use {\tt cFAC}~\cite{doi:10.1139/p07-197} to generate Dirac-Hartree-Fock wave functions. For a less precise result, which ignores the relativistic correction factor, 
one may also consult atomic-form-factor look-up tables~\cite{https://doi.org/10.1002/crat.2170280117}.

For $q\ll 10~{\rm MeV}$, the nuclear Thomson scattering amplitude, $A_\perp^{\rm N}$, is given by
\begin{equation}
    A_\perp^{\rm N}=-Z^2\frac{m_e}{m_N}\,,\quad ~A_\parallel^{\rm N}=A_\perp^{\rm N}\cos\theta\,,
\end{equation}
where $Z$ is the atomic number.
The photon can also scatter via virtual electron-positron pairs, called Delbr\"uck scattering.  This is given by
\begin{equation}
    A_\perp^{\rm D}=\alpha^2Z^2a_\perp^{\rm D}\,,~\quad A_\parallel^{\rm D}=\alpha^2Z^2a_\parallel^{\rm D}\,,
\end{equation}
where $a_\perp^{\rm D}$ and $a_\parallel^{\rm D}$ are tabulated in~\cite{FALKENBERG19921}.
Lastly, a relatively small contribution comes from the photo-nuclear absorption in the giant-resonance region. We refer to it as the nuclear resonance scattering, and it is given by~\cite{PhysRevC.23.1375,Rullhusen:1979zza,Rullhusen:1979zz,Berman:1975tt}
\begin{equation}
    A_\perp^{\rm NR}=\frac{E_\gamma^2m_e}{2\pi^2\alpha}\sigma_{-2}\,,~\quad A_\parallel^{\rm NR}=A_\perp^{\rm NR}\cos\theta\,,
\end{equation}
where
\begin{equation}
    \sigma_{-2}=0.00225\ A^{5/3}~{\rm mb/MeV}\,,
\end{equation}
with $A$ being the atomic mass.

Fig.~\ref{fig:contr} shows the total cross section for photons scattering in Si and its respective individual contributions. It is clear that the photon-electron Rayleigh scattering dominates over the other contributions for momentum transfers smaller than $100~{\rm keV}$. This is because the other processes describe photon-nucleus scatterings and the cross sections are either suppressed by $(m_e/m_N)^2$ or are higher order in $\alpha$. Thus, they only become important when $|g(q)/Z_{\rm ion}|^2$ drops below that suppression. We have confirmed that the same is true for all materials included in our analysis. Thus, the dominant process in partially coherent photon-ion scattering with energy transfers $\lesssim 100 \, \text{meV}$ is well described by the photon-electron Rayleigh scattering cross section (in agreement with~\cite{Robinson_2017}), 
\begin{equation} \label{Thompson}
  \frac{d\sigma_T}{d\Omega}({\bf q},\theta) \simeq \frac{\alpha^2}{2m_e^2}\left(1 + \cos^2{\theta}\right)|g(|{\bf q}|)|^2 \,,   
\end{equation}
which, using Eq.~\eqref{qtheta}, can be  rewritten as:
\begin{equation} \label{Thompsonp}
  \frac{d\sigma_T}{dq}({\bf q},E_{\gamma}) \simeq \frac{|{\bf q}|}{E^2_{\gamma}} \frac{\alpha^2 \pi}{m_e^2}\left(1 +\left(1-\frac{|{\bf q}|^2}{2E^2_{\gamma}}\right)^2\right)|g(|{\bf q}|)|^2 \,. 
\end{equation}
Eq.~\eqref{Thompsonp} illustrates that the photon-ion scattering cross section at leading order scales quadratically with $E_{\gamma}$ (the term $\frac{|{\bf q}|^2}{2 E^2_{\gamma}}$ is small for scattering in the forward direction when $|{\bf q}| \ll E_\gamma$). 

\section{Phonons from Photon-Ion Scattering}
\label{sec:analysis2}

Having identified the dominant photon-ion scattering process as photon-electron Rayleigh scattering, we now turn our attention to phonon production.
In condensed matter systems, electron-photon scattering can give rise to Kramer-Heisenberg polarization, which allows for the absorption and subsequent emission of photons by electrons~\cite{Kramers:1925ucm}. It is often discussed in the context of Raman scattering, and it is relevant for phonon production for photon wavelengths in the IR and visible regime $< \text{O}(\text{eV})$. For high energy photons ($> \text{keV}$), however, this process is suppressed~\cite{CROWLEY201455, doi:10.1063/5.0011416}, which can be qualitatively understood by the following argument. 
In the Kramer-Heisenberg process, the momentum of the scattered electron is transferred to the ions in the intermediate state of the second order transition processes. In our case, the photon energy is around $1~{\rm MeV}$ and there is no resonance in the material that matches this energy. Thus, the lifetime of the virtual states is too short for the creation of phonons via the Kramer-Heisenberg process.  Another way to understand this is that the photon frequency is too high to polarize the electrons in the material, and hence no phonons can be generated from polarization. 

Instead of the Kramer-Heisenberg process, phonons are generated through photon-electron Rayleigh scattering by the electrons transferring their momentum to the ions.  
This allows us to write the structure function as 

 \begin{equation}
S({\bf q}, \omega) =  \sum_f {|\langle f |e^{i {\bf q} \cdot  \hat{\bf u}} |i\rangle   |}^2 \delta(E_i - E_f - \omega)\,, \label{eq_structure_func}
 \end{equation}
where $\hat{\bf u}$ denotes the displacement of the ion and $i$ and $f$ indicate the ion's initial and final states, respectively. This partial dynamic structure factor is the same one that appears in neutron scattering experiments, and it probes the same degrees of freedom (for details, see Appendix~\ref{sec_str_fun}). For a monatomic system, it can be expressed in terms of the Debye-Waller factor $W(\bf{q})$ and the time-dependent expectation value of the ion displacement $\mathcal X({\bf q},t) = \langle {\bf q}\cdot \hat {\bf u}(0){\bf q}\cdot \hat {\bf u}(t)\rangle$~\cite{PRICE19861}, 
 \begin{equation} \label{PDSF}
S({\bf q}, \omega) =   \int_{-\infty}^{\infty} \frac{dt}{2\pi} e^{- i \omega t} e^{-2W({\bf q})} e^{\mathcal X({\bf q},t)}\,.
 \end{equation}
Both the Debye-Waller factor $W({\bf q})$ and the ion displacement $\mathcal X({\bf q},t)$ are determined by the phonon density of states (DOS) of the material $F(\omega)$, 
 \begin{equation} \label{DW}
    2W({\bf q}) = \frac{|{\bf q}|^2}{2M} \int_0^{\infty} d\omega' \frac{F(\omega')}{\omega'} (2n(\omega') +1)\,   
 \end{equation}
 \begin{equation}
    \mathcal X ({\bf q},t)=\frac{|{\bf q}|^2}{2M}\int_{-\infty}^\infty d\omega'\frac{F(|\omega'|)}{\omega'}(n(\omega')+1)e^{i\omega' t}.  
 \end{equation}
Here $n(\omega') = (e^{\omega'/T}-1)^{-1}$ is the Bose-Einstein distribution and $M$ denotes the mass of the ion. The DOS is normalized such that $\int_0^{\omega_{\text{max}}} d\omega F(\omega) = 1$\footnote{This normalization is appropriate when considering the differential cross-section per ion. If one considers the differential cross-section per unit cell the normalization differs. See Appendix \ref{sec_str_fun} for details.}. 
 
 Taylor expanding the second exponent in Eq.~\eqref{PDSF} and evaluating the time integral gives rise to single-phonon events,
 \begin{equation} \label{Sphon}
 S^{1}_{ph}({\bf q}, \omega) = e^{-2W({\bf q})} \frac{|{\bf q}|^2}{2M}\frac{F(\omega)}{\omega}(n(\omega) + 1), 
 \end{equation}
 as well as multi-phonon events that can be calculated recursively~\cite{PRICE19861},
 \begin{align}
  \label{Snphon}
   S^{n}_{ph}({\bf q}, \omega) & = &  e^{2W(|{\bf q}|)} \frac{1}{n} \int_{-\infty}^{\infty} d\omega' S^{1}_{ph}({\bf q},\omega-\omega')  \nonumber \\
  & & \times S^{n-1}_{ph}({\bf q},\omega')\,.
\end{align}

These equations highlight that the only information necessary to calculate $S({\bf q},\omega)$ in a monatomic system is the phonon density of states $F(\omega)$ for a given material. This  quantity has been directly measured in various experiments~\cite{PhysRevB.5.3151, PhysRevLett.52.644, doi:10.1063/1.331665} and also calculated from first-principles ab inito calculations~\cite{doi:10.1142/S2047684112500261, doi:10.1063/5.0017269, Petretto2018}. In this paper, for Ge, we use the experimentally measured phonon density of states at 80~K~\cite{PhysRevB.5.3151}. For the other materials (Si, GaAs, SiC, and Al$_2$O$_3$), we use the ab initio calculations of~\cite{doi:10.1142/S2047684112500261, doi:10.1063/5.0017269, Petretto2018}. Ab initio calculations are expected to match exactly with the phonon density of states at 0~K. For materials considered in this paper, the phonon density of states is not expected to vary significantly with temperature, at least up to temperatures as high as room temperature. The ab initio calculations have also been found to match well with experimental data at room temperature~\cite{doi:10.1063/1.1484241}. This validates our choice for the phonon density of states for the materials considered here. 

An advantage of using the ab initio calculations is the availability of the partial phonon density of states (pDOS) $F_i(\omega)$ for composite materials. The pDOS appropriately weights how much a given ion contributes to a phonon excitation such that $ \sum_d F_d(\omega) = F(\omega)$, allowing us to linearly decompose the calculation into the contribution of the individual ions $d$ as 
\begin{equation} \label{sumrate}
\frac{dR_{\text{total}}}{d\omega} (\omega) = \sum_d  \frac{dR_{d}}{d\omega} (\omega)\,.  
\end{equation}
Here in the calculation of $\frac{dR_d}{d\omega}$ we use the mass $M_{d}$, modified form factor $g_d(\omega),$\footnote{Using the modified form factor of a given atom versus the ion has negligible effects.} abundance in material $N_{T;d}$, as well as the pDOS $F_d(\omega)$ of the appropriate ion $d$. We also assume $n(\omega) \approx 0$, since we expect these experiments to take place at low temperatures to reduce thermal noise. A more detailed derivation of the dynamic structure function in diatomic materials is provided in Appendix~\ref{sec_str_fun}. For a material such as ${\rm Al}_2{\rm O}_3$, which does not have a cubic symmetry, the use of our formalism may introduce an uncertainty that becomes larger for larger multi-phonon multiplicity $n$ in Eq.~(\ref{Snphon}).   
In this work, we calculate up to $n = 6$, which is sufficient to capture the dominant contributions to the energy range $\omega\sim 1~{\rm meV}-100~{\rm meV}$.

 \begin{figure}
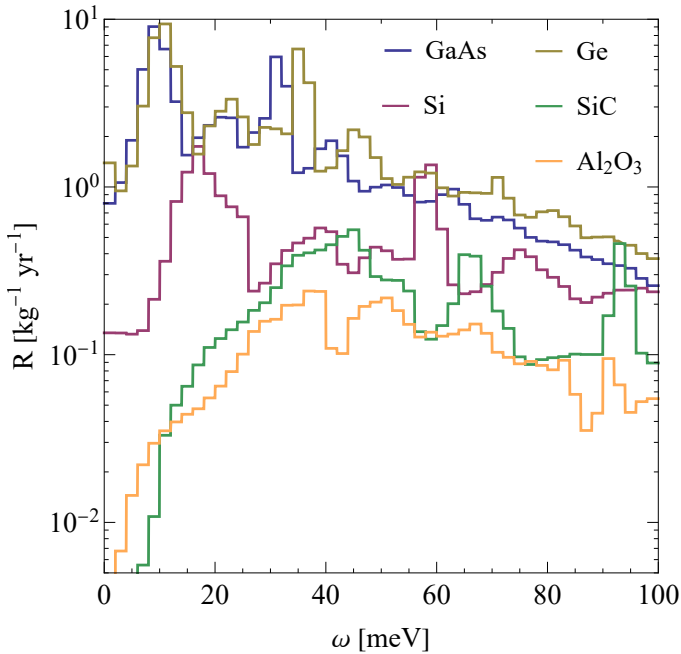

    \adjustimage{width=0.5\textwidth,center}{"result_plot6".pdf}
    \vspace{-0.7cm}
    \caption{Expected phonon spectrum from high-energy background photon scattering in silicon, germanium, gallium arsenide, silicon carbide, and sapphire.  We assume a high-energy photon background flux that creates a Compton-scatter background rate of $0.042$~events/kg/day/keV in germanium at $\mathcal{O}(10~\text{keV})$ energies (see Appendix~\ref{subsec:pb}). The displayed rates include  multi-phonon contributions up to $n =6$. The bin width corresponds to $2$~meV.}
     \label{fig:background_rates}
\end{figure} 

\section{Results}\label{sec:results} 

Following the formulation in Sec.~\ref{sec:analysis2}, we calculate and show in Fig.~\ref{fig:background_rates} the expected total phonon background rate and spectrum for Si, Ge, GaAs, SiC, and $\text{Al}_2\text{O}_3$ targets.  We assume a high-energy photon background flux that creates a Compton-scatter background rate of $0.042$~events/kg/day/keV in germanium at $\mathcal{O}(10~\text{keV})$ energies as achieved by {\it e.g.} EDELWEISS~\cite{EDELWEISS:2018tde} and SuperCDMS SNOLAB~\cite{SuperCDMS:2016wui}.  We include contributions from several distinct photon energies.  For concreteness, we take a photon background spectrum (photon energies $E_{\gamma_i}$ and densities $n_{\gamma_i}$) as measured and simulated for a Ge target in a well-shielded environment by the EDELWEISS Collaboration in~\cite{EDELWEISS:2018tde}, which we discuss in Appendix~\ref{subsec:pb} (see Table~\ref{tab:photondensities}). However, as discussed in Sec.~\ref{sec:analysis2} and Appendix~\ref{subsec:pb}, the differential cross section and thus the shape of the phonon spectrum (see Fig.~\ref{fig:comparison}) does not depend sensitively on the choice of photon energies, so a different choice of photon energies would not qualitatively affect our results.  In Appendix~\ref{sec_str_fun} (see Fig.~\ref{fig:multiphonon}), we compare the single vs. multi-phonon contributions. 

From Fig.~\ref{fig:background_rates}, we learn that numerous low energy depositions that can mimic DM signals are expected, even in a  well-shielded environment.  The low-energy background event rates are consistent with those in~\cite{Robinson_2017}.  For example, in silicon (germanium) we expect $N_{bk}  = 23$ (93) total background events in the energy range $1-100$~meV. 
This affects the number of events needed to claim a DM detection and also affects the expected $2\sigma$-sensitivities.  
These backgrounds can be reduced by improving the passive shielding, further reducing radioactive impurities in detector materials, and/or by having an active veto surround the target.  An active veto would search for Compton scatters of the high-energy photons in coincidence with a low-energy coherent scatter that creates a single- or multi-phonon event in the target. In Appendix~\ref{app:veto}, we estimate the photon-veto efficiencies in the detector without an additional active veto surrounding the target ({\it i.e.}, we estimate how often a high-energy photon will Compton scatter or be absorbed in the detector target material), and find that a large fraction of events cannot be vetoed with the detector target itself.


 \begin{figure*}[th!]
    \adjustimage{width=1\textwidth,center}{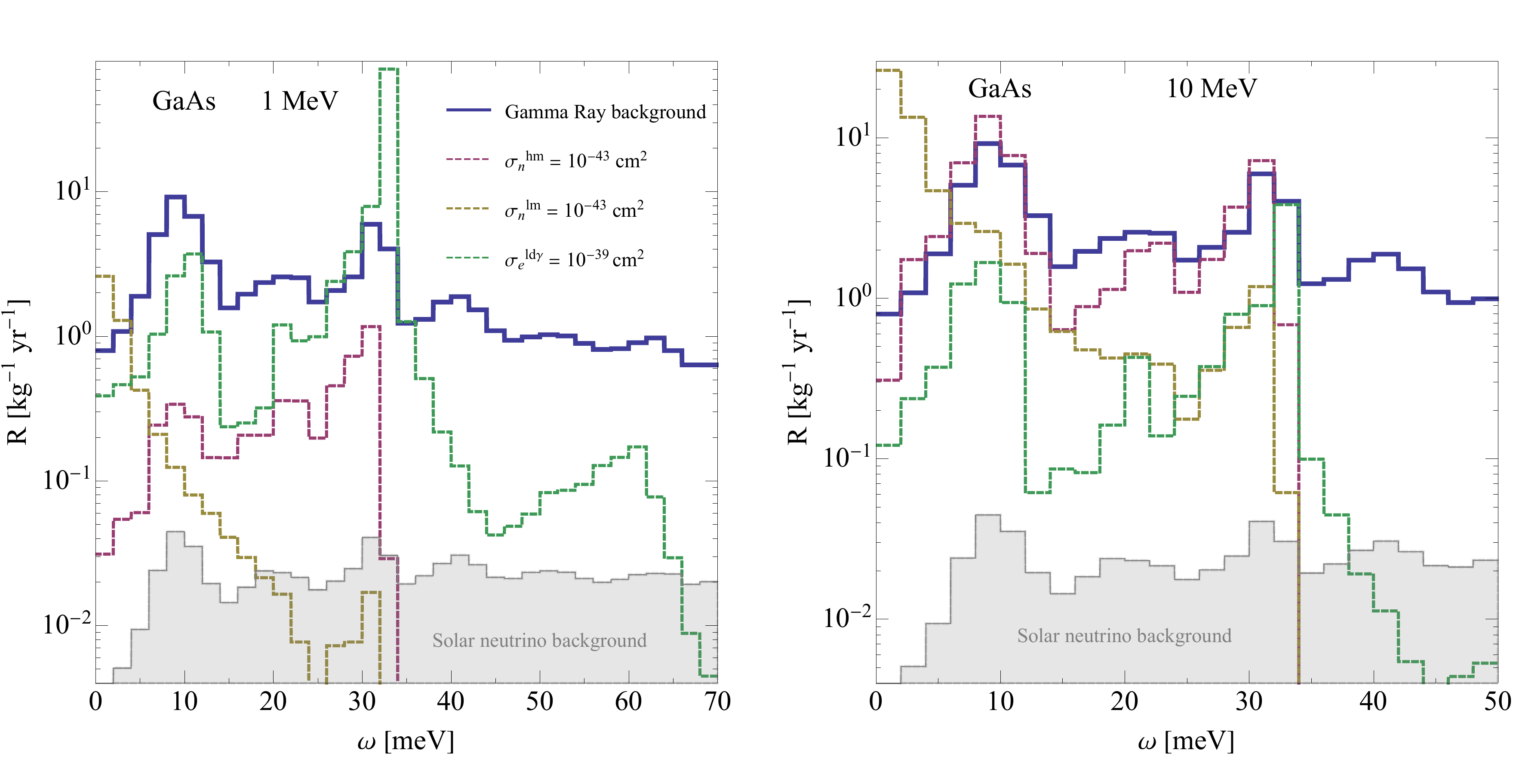}
  \vspace{-0.7cm}
    \caption{Expected phonon spectrum in GaAs from high-energy background photon scattering compared with the phonon spectrum generated by three different DM candidates (dashed) for DM masses of $1\, \text{MeV}$ on the left, and $10\,\text{MeV}$ on the right.  We assume a high-energy photon background flux that creates a Compton-scatter background rate of $0.042$~events/kg/day/keV in germanium at $\mathcal{O}(10~\text{keV})$ energies (see Appendix~\ref{subsec:pb} for details). The bin width corresponds to $2$~meV. 
    The three models shown correspond to DM coupling to nucleons via a heavy (hm, purple) and light (lm, beige) scalar mediator with a cross section of $\sigma_n = 10^{-43}\text{cm}^{2}$ \cite{Trickle_2020}, as well as a light dark photon mediator model (ld$\gamma$, green) that couples to the atomic constituents proportionally to the Standard Model photon \cite{Knapen:2021bwg}. Shown for comparison in gray is the solar neutrino background from coherent neutrino-nucleus scattering (see Appendix~\ref{subsec:solar_neutrin} for details).  
     }  
     \label{fig:signal_rates}
\end{figure*} 

 We now compare the phonon backgrounds from photon scattering in the target with the expected DM signal shape.  The DM signal strongly depends on the DM model. Whether the DM couples to the nucleus directly or to electrons (or both) can modify the dynamic structure function $S(\bf{q}, \omega)$, which leaves an imprint on the signal spectrum. Direct couplings to only the nucleus lead to the same $S(\bf{q},\omega)$ as the partially coherent photon-ion scattering background due to the phonon excitation being facilitated by a net momentum transfer to the ion. DM couplings to electrons can, however, significantly modify $S(\bf{q},\omega)$ by, for example, polarizing the material. Integrating over the DM velocity distribution additionally places kinematic constraints on which energy deposits are accessible with a given DM mass. Generally, phonon excitations are most relevant for DM masses below $ m_{\text{DM}} \sim 100$~MeV. Lastly, light mediator models get a cross-section enhancement at small $q$, leading to a relative enhancement for smaller energy deposits. 
 
 We write the differential DM rate as 
 \begin{equation}
    \frac{dR}{d\omega}(\omega) = N_T \frac{\rho_{\rm DM}}{m_{\text{DM}}} \int  dv^3 \int dq  f_{\text{DM}} v q \frac{d\sigma_{\rm DM}}{dq} S_{\text{DM}}\,. 
 \end{equation}
 Here $v$ is the DM velocity and $f_{\text{DM}}$ the DM velocity distribution typically taken to be a truncated Boltzmann-distribution \cite{LEWIN199687}. The DM mass $m_{\text{DM}}$, cross-section $\frac{d\sigma_{\text{DM}}}{dq}$, and triggered material transition channel determine the final signal spectrum. 
 
 In Fig.~\ref{fig:signal_rates}, we contrast the coherent photon background for GaAs with the signal spectrum of three distinct DM models and two DM masses, $m_{\text{DM}} = 1 \, \text{MeV}$ ({\it left}) and $m_{\text{DM}} = 10 \, \text{MeV}$ ({\it right}) . 
 These models are DM that interacts with either a light or heavy scalar mediator that couples equally to protons and neutrons~\cite{Trickle:2019nya} (denoted as `lm' and `hm', respectively), as well as DM interacting via a light dark photon mediator (denoted `ld$\gamma$'), which couples to the atomic constituents in the same way as the Standard Model photon.
For Fig.~\ref{fig:signal_rates}, we extract the signal for the scalar mediators directly from Fig.~5 of \cite{Trickle:2019nya}, and use a combination of the \href{https://demo-phonon-web-app.herokuapp.com/}{\tt Dark Matter-Single Phonon Interaction Rate Calculator} \cite{Trickle:2019nya, Griffin:2019mvc, Trickle:2020oki, Coskuner:2021qxo}  and the \href{https://github.com/tongylin/DarkELF} {\tt DarkELF} package~\cite{Knapen:2021bwg} to calculate DM interacting with an ultralight (or massless) dark photon mediator. We use the former to estimate the rate below the single phonon threshold ($\omega_{\text{max}} = 34 \text{ meV}$ for GaAs), and the latter to estimate the rate above it. DarkELF utilizes the energy-loss-function  $\text{Im}\left[\frac{-1}{\epsilon(\omega)}\right]$~\cite{Hochberg:2021pkt,Knapen:2021run} in the zero momentum limit, which for DM above 1 MeV gives a good estimate of the rate above the optical single phonon threshold, but does not include acoustic mode contributions below it.\footnote{We shifted the \href{https://github.com/tongylin/DarkELF} {\tt DarkELF} spectrum such that its peak on the optical resonance matches with the highest optical phonon frequency of the GaAs phonon DOS utilized in the background calculation as well as the  \href{https://demo-phonon-web-app.herokuapp.com/}{\tt Dark Matter-Single Phonon Interaction Rate Calculator}.}

 From Fig.~\ref{fig:signal_rates}, we see that the heavy scalar mediator model signal (in red) most closely resembles the background spectrum, which is especially noticeable for  heavier DM. Small differences are due to kinematics as well as the inclusion of multi-phonon contributions in our background calculation which were not included in the signal calculation. For both DM masses of 1~MeV and 10~MeV for the heavy mediator model, significant contributions from Umklapp processes arise, since the DM carries momentum that extends beyond one Brillouin zone (see~\cite{Trickle_2020} for details). The light scalar mediator model 
 peaks at small momentum transfers due to the light mediator enhancement $\frac{d\sigma_{\text{DM}}}{dq} \propto  \frac{q_0^4}{q^4}$, with $q_0 \sim \alpha m_e$, which leads to a relative enhancement of small energy deposits $\omega$, as is clearly noticeable in the beige curves of Fig.~\ref{fig:signal_rates}. Lighter masses have smaller overall rates for the scalar mediator models due to typical momenta being smaller (and the resulting absence of Umklapp contributions). For the case of a light dark photon mediator, the signal grows towards smaller masses down to about $m_{\rm DM} \approx 3\,  \text{keV}$, when the rates become kinematically suppressed and the signal carries very little energy. The enhancement towards lighter DM masses above the kinematic threshold comes from the larger DM number density for lighter DM without a momentum-dependent suppression in the energy-loss-function based dynamic structure function. Interestingly, for a light dark photon mediator, the energy deposition peaks strongly at the optical phonon resonance ($34 \,\text{meV}$) in GaAs. This behavior is the same for other polar materials. For non-polar materials (Si, Ge), multi-phonons dominate the scattering process, leading to energy deposits above the single-phonon threshold. The spectral shape can be used to discriminate between the background and the light dark photon mediator model, since the photon-ion scattering background is not as peaked towards the optical phonon resonance or multi-phonons. Qualitatively this happens due to the additional Debye-Waller factors $\sim \left(\frac{q^2}{2M \overline{\omega}_{\text{ph}}} \right)^n$ being present for higher-order terms, which are always smaller than 1 due to the modified form factor $g(q)$ only being sizable for small $q$ ($\lesssim 100$~keV). Here $\overline{\omega}_{\text{ph}}$ denotes the average phonon energy. For both scalar mediator models, discriminating between the background and signal spectra proves more challenging due to their similarity, which relates to the same partial dynamic structure factor $S(\bf{q},\omega)$ entering the background and signal rate calculations. The properties of these differences depend on the specific DM and mediator masses, where perhaps the light mediator can be distinguished based on the relative low energy deposition enhancement, which could be exploited if detectors have high energy resolution and extended sensitivity to small energy depositions. The solar neutrino background (see Appendix~\ref{subsec:solar_neutrin}  for details) is suppressed by a factor of $10^2$ relative to the gamma-ray background. 
 
 We make the calculation of our background spectra publicly available on \href{https://github.com/KBerghaus/cgb}{\tt Github}, where we provide look-up tables for the phonon densities and the modified atomic form factors as well as an easily usable mathematica notebook.

\section{Conclusions}
\label{sec:disc}

In this work, we presented the phonon background generated by high-energy photons scattering in the solid-state targets Si, Ge, GaAs, SiC, and $\text{Al}_2\text{O}_3$. These target materials have all been proposed for sub-GeV DM searches.  We discuss different contributions to this background and show how to calculate the expected background rate given current shielding capabilities in existing experiments, using  EDELWEISS-III~\cite{EDELWEISS:2018tde} as our benchmark. We find that using the background-substracted photoabsorption peaks from various high-energy photons ranging from $100 \,\text{keV}$ to $10 \, \text{MeV}$ produces 0.042~events/day/kg/keV assuming a flat Compton background in germanium. Using these photon densities, we calculated the expected phonon background for the materials listed above. We find that the background spectrum has many similarities with those of scalar mediator models with DM coupling to nucleons due to the shared dynamic structure function. Additionally, light scalar mediators predict a signal enhancement at smaller energy depositions, whereas DM interacting with light dark photons have a phonon spectrum that peaks at larger energy deposits due to its ability to polarize the material, which leads to a qualitatively different dynamic structure function. These differences may be exploited in signal-versus-background discrimination.  
We showed that the detector target itself is not sufficient to veto the high-energy background photons, and hence an active veto is needed to allow for a background-free DM search for large exposures. 
We make our calculation publicly available on \href{https://github.com/KBerghaus/phonon_background}{\tt Github}~\cite{BerghausGithub}, allowing the community to easily include these backgrounds in future theoretical and experimental investigations.

\begin{acknowledgments}
We thank Cyrus Dreyer, Marivi Fern\'andez-Serra, Sin\'ead Griffin, Matt Pyle, Bjoern Penning, Tongyan Lin, Alan Robinson, and Tanner Trickle for useful discussions. We also thank Matt Pyle for comments on the manuscript and for the suggestion to show the single- and multi-phonon backgrounds from coherent neutrino-nucleus scattering due to solar neutrinos. We are also grateful to Lin-Fan Zhu and Michael Walter for their comments on the relevance of Raman-activity for MeV-photons. KB acknowledges the support of NSF grant PHYS-1915093. RE acknowledges support from DoE Grant DE-SC0009854, Simons Investigator in Physics Award 623940, and the US-Israel Binational Science Foundation Grant No.~2020220.  The work of YH  is supported by the Israel Science Foundation (grant No. 1112/17), by the Binational Science Foundation (grant No. 2016155), by the I-CORE Program of the Planning Budgeting Committee (grant No. 1937/12), and by the Azrieli Foundation. The work of YS is supported by the I-CORE Program of the Planning Budgeting Committee (grant No. 1937/12). MS acknowledges support from DoE Grants DE-SC0009854, DE-SC0009919, and DE-SC0022104. 

\end{acknowledgments}

\appendix 

\section{Structure Function}
\label{sec_str_fun}
In this appendix, we formulate the inelastic scattering cross section and derive Eq.~\eqref{eq_structure_func} following~\cite{schober_2014}. 
We consider a crystal consists of $n$ atoms in a primitive cell. The position of an atom is expressed as
\begin{equation}
    {\bf R}_{d{\boldsymbol \ell}}={\boldsymbol \ell}+{\bf r}_d,
\end{equation}
where ${\boldsymbol \ell}$ points a primitive cell and ${\bf r}_d$ is the position of atom $d$ in a cell. 

We assume that the atom-photon interaction potential, $\hat V_I$, has matrix elements of
\begin{align}
    &\langle {\bf k}_f,\lambda_f,\mathcal A_f^{d{\boldsymbol \ell}}|\hat V_I| {\bf k}_i,\lambda_i,\mathcal A_i^{d{\boldsymbol \ell}}\rangle\nonumber\\
    &=\frac{e^2}{2m_e}\frac{1}{\sqrt{|{\bf k}_i||{\bf k}_f|}}\langle \mathcal A_f^{d{\boldsymbol \ell}}|e^{i({\bf k}_i-{\bf k}_f)\cdot\hat{\bf R}_{d{\boldsymbol \ell}}}|\mathcal A_i^{d{\boldsymbol \ell}}\rangle\nonumber\\
    &\hspace{3ex}\times\left[a_1^d({\boldsymbol \epsilon}^*_{\lambda_f{\bf k}_f}\cdot{\boldsymbol \epsilon}_{\lambda_i{\bf k}_i})+a_2^d({\boldsymbol \epsilon}^*_{\lambda_f{\bf k}_f}\cdot {\bf k}_i)({\boldsymbol \epsilon}_{\lambda_i{\bf k}_i}\cdot {\bf k}_f)\right],
\end{align}
where ${\bf k}_i(\lambda_i)$ and ${\bf k}_f(\lambda_f)$ are the initial and the final photon momenta (helicities), and $\mathcal A_i^{d{\boldsymbol \ell}}$ and $\mathcal A_f^{d{\boldsymbol \ell}}$ are the initial and the final atomic states at $R_{d{\boldsymbol \ell}}$.
The coefficient, $a_i^d$, is a function of ${\bf k}_i$ and ${\bf k}_f$. We work in the Coulomb gauge, {\it i.e.} ${\boldsymbol \epsilon}_{\lambda,{\bf k}}\cdot {\bf k}=0$. The position operator, $\hat{\bf R}_{d{\boldsymbol \ell}}$, operates on the whole atom and does not affect the internal states.
Here, $\hat V_I$ should be understood as an effective interaction potential including the contributions of the second order scattering processes. 

Going back to a crystal, we consider the transition from initial state $\mathcal S_i$ to final state $\mathcal S_f$.
The distribution of the initial states is $p(\mathcal S_i)$.
From Fermi's golden rule, the inclusive differential cross section is given by
\begin{align}
    \frac{d^2\sigma}{d\Omega d|{\bf k}_f|}&=\frac{|{\bf k}_f|^2}{(2\pi)^3}\frac12\sum_{\lambda_i\lambda_f}\sum_{\mathcal S_i\mathcal S_f}p(\mathcal S_i)\nonumber\\
    &\hspace{3ex}\times|\langle {\bf k}_f,\lambda_f,\mathcal S_f|\hat V_I| {\bf k}_i,\lambda_i,\mathcal S_i\rangle|^2\nonumber\\
    &\hspace{3ex}\times2\pi\delta(E^f_{\mathcal S}+|{\bf k}_f|-E^i_{\mathcal S}-|{\bf k}_i|),
\end{align}
where $E^i_{\mathcal S}$ and $E^f_{\mathcal S}$ are the initial and final state energy of the target.

Working in a basis where the photon has a linear polarization that is perpendicular ($\lambda=\perp$) or parallel ($\lambda=\parallel$) to the scattering plane,\footnote{The additional $|{\bf k}_f|/|{\bf k}_i|$ factor appearing in the Klein-Nishina formula originates from the delta function.}
\begin{align}
    \frac{d^2\sigma}{d\Omega d|{\bf k}_f|}&=\frac{\alpha^2}{4\pi m_e^2}\frac{|{\bf k}_f|}{|{\bf k}_i|}\sum_{\lambda}\sum_{\mathcal S_i\mathcal S_f}p(\mathcal S_i)\nonumber\\
    &\hspace{3ex}\times\left|\sum_{d{\boldsymbol \ell}}A_\lambda^d\langle \mathcal S_f|e^{i({\bf k}_i-{\bf k}_f)\cdot \hat{\bf R}_{d{\boldsymbol \ell}}}|\mathcal S_i\rangle\right|^2\nonumber\\
    &\hspace{3ex}\times2\pi\delta(E^f_{\mathcal S}+|{\bf k}_f|-E^i_{\mathcal S}-|{\bf k}_i|),
\end{align}
where
\begin{align}
    A_\perp^d=a_1^d,~A_\parallel^d=a_1^d\cos\theta-a_2^d|{\bf k}_i||{\bf k}_f|\sin^2\theta.
\end{align}

Using
\begin{equation}
    \delta(E)=\int \frac{dt}{2\pi}e^{-iEt},
\end{equation}
and summing over $\mathcal S_f$, we obtain
\begin{align}
    \frac{d^2\sigma}{d\Omega d|{\bf k}_f|}&=\frac{\alpha^2}{2 m_e^2}\frac{|{\bf k}_f|}{|{\bf k}_i|}\sum_{\lambda}\sum_{\mathcal S_i}p(\mathcal S_i)\nonumber\\
    &\hspace{3ex}\times\sum_{d{\boldsymbol \ell}}\sum_{d'{\boldsymbol \ell}'}A_\lambda^dA_\lambda^{d'*}\int \frac{dt}{2\pi} e^{-i\omega t}\nonumber\\
    &\hspace{3ex}\times\langle \mathcal S_i|e^{-i{\bf q}\cdot \hat{\bf R}_{d'{\boldsymbol \ell}'}(0)}e^{i{\bf q}\cdot \hat{\bf R}_{d{\boldsymbol \ell}}(t)}|\mathcal S_i\rangle\,,
\end{align}
where ${\bf q}={\bf k}_i-{\bf k}_f$, $\omega =|{\bf k}_i|-|{\bf k}_f|$ and
\begin{equation}
    e^{i{\bf q}\cdot \hat{\bf R}_{d{\boldsymbol \ell}}(t)}=e^{i\hat Ht}e^{i{\bf q}\cdot \hat{\bf R}_{d{\boldsymbol \ell}}}e^{-i\hat Ht}\,.
\end{equation}

\begin{figure*}[h]
    \adjustimage{width=\textwidth,center}{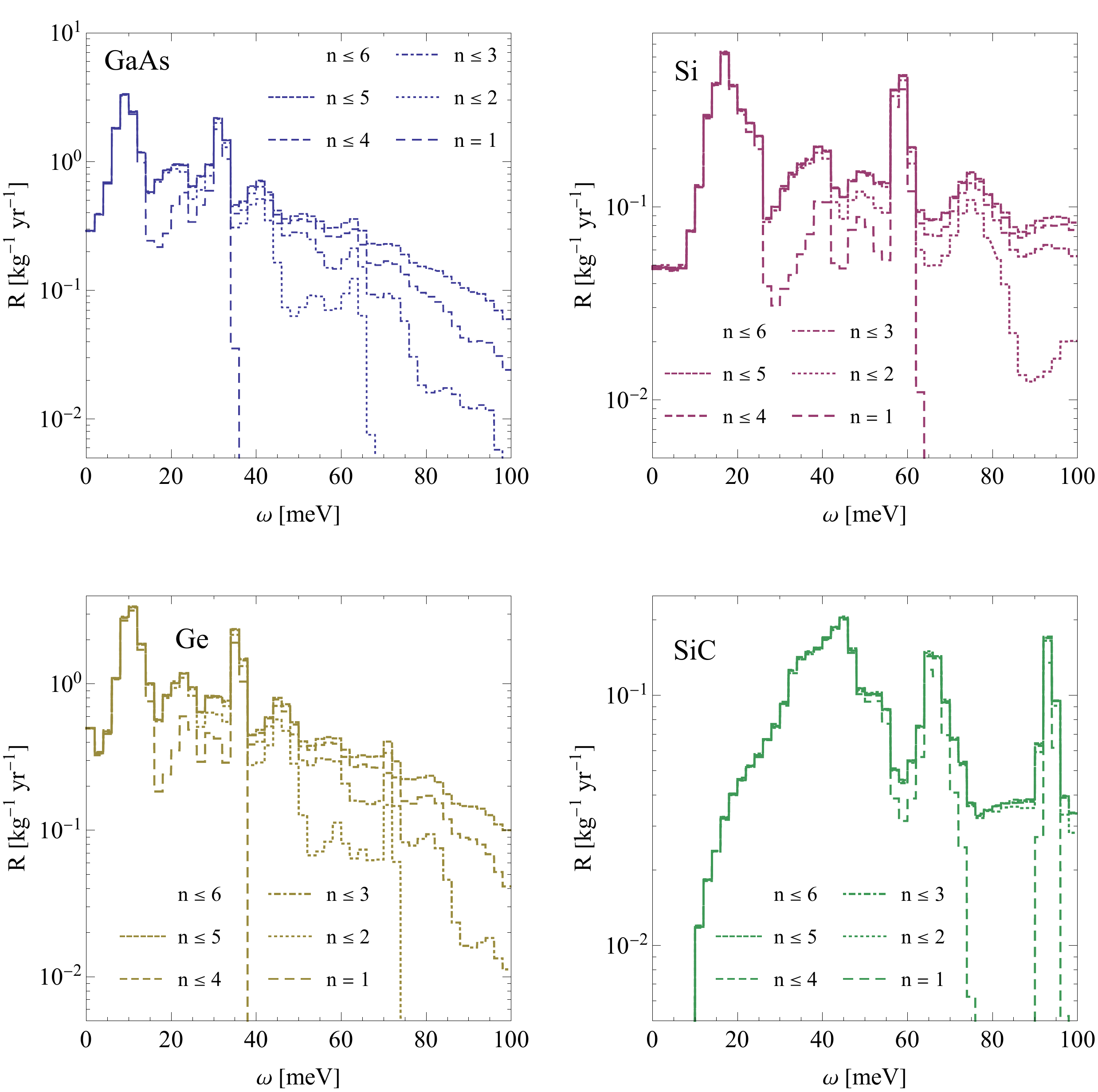}
    \vspace{-0.7cm}
    \caption{The cumulative multi-phonon contributions to the total rate up to $n\leq 6$ for GaAs, Si, Ge, and SiC (see Eq.~\eqref{Sd} for details). Shown spectra correspond to background rates assuming a single photon energy $E_{\gamma} = 1.4 61$ MeV ($n_{\gamma} = 4.7 \times 10^{-16}\,  \text{cm}^{-3}$). Only including single phonons (dashed, large spacing) captures the peak of the spectra well below the cutoff of the phonon density 
    $\omega_{\text{max}}$ for a given material which corresponds to $0.35\,\text{meV}$, $0.68 \, \text{meV}$, $0.38\, \text{meV}$, and $0.12 \, \text{meV}$ for GaAs, Si, Ge, and SiC, respectively. However, multi-phonons can give sizeable contributions off peaks even below $\omega_{\text{max}}$. 
    Above the cutoff the spectrum is dominated by higher order multi-phonons where each multi-phonon can contribute up to a maximum energy of 
    $\omega = n \omega_{\text{max}}$. 
    Higher orders are suppressed by additional Debye-Waller factors which together with the atomic form factors being only sizeable at small $q$ ($\lesssim$ 100~keV) leads to a convergent expansion.\label{fig:multiphonon}}
\end{figure*}

We define
\begin{equation}
    \langle \hat A\rangle=\sum_{\mathcal S}p(\mathcal S)\langle\mathcal S|\hat A|\mathcal S\rangle\,.
\end{equation}
Then, for $[\hat A,\hat B]=\rm constant$, we have
\begin{equation}
    \langle\exp\hat A\exp\hat B\rangle=\exp\left(\frac12\langle\hat A^2+\hat B^2+2\hat A\hat B\rangle\right)\,.
\end{equation}
Thus,
\begin{align}
    \frac{d^2\sigma}{d\Omega d|{\bf k}_f|}&=\frac{\alpha^2}{2 m_e^2}\frac{|{\bf k}_f|}{|{\bf k}_i|}\sum_{\lambda}\sum_{d{\boldsymbol \ell}}\sum_{d'{\boldsymbol \ell}'}A_\lambda^{d'*}A_\lambda^d\nonumber\\
    &\hspace{3ex}\times e^{-i{\bf q}\cdot(\bar{\bf R}_{d'{\boldsymbol \ell}'}-\bar{\bf R}_{d{\boldsymbol \ell}})}e^{-W_{d'}({\bf q})}e^{-W_{d}({\bf q})}\nonumber\\
    &\hspace{3ex}\times\int \frac{dt}{2\pi} e^{-i\omega t}e^{\langle {\bf q}\cdot \hat{\bf u}_{d'{\boldsymbol \ell}'}(0){\bf q}\cdot \hat{\bf u}_{d{\boldsymbol \ell}}(t)\rangle}\,,
\end{align}
where
\begin{align}
    \hat{\bf R}_{d{\boldsymbol \ell}}&=\bar{\bf R}_{d{\boldsymbol \ell}}+\hat{\bf u}_{d{\boldsymbol \ell}}\,,\\
    W_d({\bf q})&=\frac12\langle [{\bf q}\cdot \hat{\bf u}_{d{\boldsymbol \ell}}(t)]^2\rangle\,.
\end{align}
Here, $\bar{\bf R}_{d{\boldsymbol \ell}}$ is the stationary point, and $W_d({\bf q})$ is called the Debye-Waller factor, which is independent of $t$ and ${\boldsymbol \ell}$.

Since we are interested in the momentum transfer that is much larger than the atomic scale, $q\gg 1~{\rm keV}$, we ignore the coherent scattering over atoms and consider $d'=d$ and ${\boldsymbol \ell}'={\boldsymbol \ell}$. We have
\begin{align}
    \frac{d^2\sigma}{d\Omega d|{\bf k}_f|}&= \frac{|{\bf k}_f|}{|{\bf k}_i|}\sum_{d}\frac{d\sigma_d}{d\Omega}S_d({\bf q},\omega )\,,\label{eq_xsec_full}
\end{align}
where
\begin{align}
\frac{d\sigma_d}{d\Omega}&=\frac{\alpha^2}{2 m_e^2}\sum_{\lambda}|A_\lambda^d|^2\,,\\
S_d({\bf q},\omega )&=e^{-2W_{d}({\bf q})}\int \frac{dt}{2\pi} e^{-i\omega t}e^{\mathcal X_d({\bf q},t)}\,,\\
    \mathcal X_d({\bf q},t)&=\langle {\bf q}\cdot \hat{\bf u}_{d{\boldsymbol \ell}}(0){\bf q}\cdot \hat{\bf u}_{d{\boldsymbol \ell}}(t)\rangle\,.
\end{align}
Here, we divided by the the number of primitive cells, $N$, to make it the cross section per  primitive cell.

In the harmonic oscillator approximation, $\hat{u}_{d{\boldsymbol \ell}}$ can be written as
\begin{equation}
    [\hat{u}_{d{\boldsymbol \ell}}]_\alpha=\frac{1}{\sqrt{N}}\sum_{j,{\bf q}}\frac{1}{\sqrt{2\omega_{j,{\bf q}}M_d}}[\hat{a}_{j,{\bf q}}+\hat{a}_{j,-{\bf q}}^\dagger]U^{d,\alpha}_{j,{\bf q}}e^{i{\bf q}{\boldsymbol \ell}},
\end{equation}
where $\hat a_{j,{\bf q}}$ and $\hat a_{j,{\bf q}}^\dagger$ are the annihilation and the creation operators for the phonon with energy $\omega_{j,{\bf q}}$. Index $\alpha=1\dots3$ is the spacial index. The oscillation modes are diagonalized and labeled with branch $j$ and momentum $q$ in the first Brillouin zone. The diagonalization matrix is denoted as $U_{j,{\bf q}}^{d,\alpha}$.

The Debye-Waller factor becomes
\begin{equation}
    W_d({\bf q})=\frac{1}{4NM_d}\sum_{j{\bf q}'}\frac{\left|\sum_\alpha q_\alpha U^{d\alpha}_{j{\bf q}'}\right|^2}{\omega_{j{\bf q}'}}(2n_{j{\bf q}'}+1)\,,
\end{equation}
where
\begin{equation}
    n_{j{\bf q}}=\langle a^\dagger_{j{\bf q}}a_{j{\bf q}}\rangle\,.
\end{equation}
As for the time-dependent factor,
\begin{align}
    \mathcal X_d({\bf q},t)&=\frac{1}{2M_dN}\sum_{j{\bf q}'}\frac{\left|\sum_\alpha q_\alpha U^{d\alpha}_{j{\bf q}'}\right|^2}{\omega_{j{\bf q}'}}\nonumber\\
    &\hspace{3ex}\times\left[(n_{j{\bf q}'}+1)e^{i\omega_{j{\bf q}'} t}+n_{j{\bf q}'}e^{-i\omega_{j{\bf q}'} t}\right]\,.
\end{align}

We further assume a cubic system. Since $W_d$ and $\mathcal X_d$ should be independent under the cubic transformation, {\it e.g.} $(q_1,q_2,q_3)\to(q_1,-q_2,-q_3)$, the cross terms of $\left|\sum_\alpha q_\alpha U^{d\alpha}_{j{\bf q}'}\right|^2$ disappear after summing over $j$ and ${\bf q}'$. In addition, $\left|U^{d\alpha}_{j{\bf q}'}\right|^2$ does not depend on $\alpha$ due to the cubic symmetry. Thus, the formulae are simplified as
\begin{equation}
    W_d({\bf q})=\frac{|{\bf q}|^2}{4M_d}\int_0^\infty d\omega\frac{F_d(\omega)}{\omega}(2n(\omega)+1)\,,
\end{equation}
and
\begin{align}
    \mathcal X_d({\bf q},t)&=\frac{|{\bf q}|^2}{2M_d}\int_0^\infty d\omega\frac{F_d(\omega)}{\omega}\nonumber\\
    &\hspace{3ex}\times\left[(n(\omega)+1)e^{i\omega t}+n(\omega)e^{-i\omega t}\right]\,,
\end{align}
where
\begin{equation}
    F_d(\omega)=\frac{1}{3N}\sum_{j{\bf q}}\sum_\alpha\left|U^{d\alpha}_{j{\bf q}}\right|^2\delta(\omega-\omega_{j{\bf q}})\,.
\end{equation}
Here, $F_d(\omega)$ is called the partial degrees of freedom and
\begin{equation}
    \int_0^\infty d\omega F_d(\omega)=1\,.
\end{equation}
For temperature $T$, we have
\begin{equation}
    n(\omega)+1=\frac{1}{e^{\omega/T}-1}+1=-n(-\omega)\,.
\end{equation}
Thus, we can rewrite $\mathcal X_d({\bf q},t)$ as
\begin{align}
    \mathcal X_d({\bf q},t)&=\frac{|{\bf q}|^2}{2M_d}\int_{-\infty}^\infty d\omega\frac{F_d(|\omega|)}{\omega}(n(\omega)+1)e^{i\omega t}\,.
\end{align}

The time integral in Eq.~\eqref{eq_xsec_full} can be evaluated by expanding $\mathcal X_d({\bf q},t)$,
\begin{align} \label{Sd}
S_d({\bf q},\omega )&=e^{-2W_{d}({\bf q})}\sum_{n=0}^\infty\frac{1}{n!}\left(\frac{|{\bf q}|^2}{2M_d}\right)^n\nonumber\\
    &\hspace{3ex}\times\left[\prod_{i=1}^n\int_{-\infty}^\infty d\omega_i\frac{F_d(|\omega_i|)}{\omega_i}(n(\omega_i)+1)\right]\nonumber\\
    &\hspace{3ex}\times\delta\left(\omega -\sum_{i=1}^n\omega_i\right)\nonumber\\
    &=\sum_{n=0}^\infty S_d^n({\bf q},\omega )\,.
\end{align}
Here, $n=0$ corresponds to the elastic scattering, where the momentum transfer turns into the momentum of the whole material without creating phonons. The terms with $n>0$ change the phonon number and describe the inelastic scattering with $n$ phonon creation/annihilation. Fig.~\ref{fig:multiphonon} shows a breakdown of the background rates based on how many multi-phonon terms are included in the expansion for a selection of the materials discussed in this work.

\section{Expected High-Energy Photon Background Flux}
\label{subsec:pb}
The magnitude of the photon-ion scattering background depends on the flux of high-energy photons incident on the detector, which in turn depends on the particular detector setup, geometry, shielding, and the radioactive impurity concentrations of the detector components.  In addition, an active veto can render the background less important, since the high-energy photon may Compton scatter in the target or an active veto detector either before or after creating a phonon signal in the target.  To estimate this background, we consider the photon flux observed or expected in current well-shielded experiments. For concreteness in this paper, we model the input photon flux based on the simulated background of the EDELWEISS-III detector~\cite{EDELWEISS:2018tde}. For photon energies below $\sim$1~MeV,  we take the continuum-subtracted rate in~\cite{EDELWEISS:2018tde} and,  assuming the peaks in the rate to be produced by photoabsorption, we calculate the photon densities that the detector encountered. For photon energies above $\sim$1~MeV, we use the electron recoil rates shown in~\cite{Scorza:2015vla}, and manually subtract the continuum background to isolate the peaks and calculate the respective photon densities. The photon densities extracted with this method are given in Table~\ref{tab:photondensities}.

\begin{table}[!t]
 \begin{tabular}{|c|c|c|}
 \hline
  $E_{\gamma}$ [MeV] &  Source   & $n_{\gamma} [\times 10^{-18}~\text{cm}^{-3}$] \\
 \hline
 0.143 & $^{235}\text{U}$ & 0.176 \\
 0.163 & Unidentified & 0.143 \\
 0.185 & $^{235}\text{U}$, $^{226}\text{Ra}$ & 1.68 \\
0.208 &  $^{232}\text{Th}$ & 1.39 \\
 0.238 & $^{232}\text{Th}$, $^{226}\text{Ra}$ & 16.65 \\
0.269 & Unidentified & 1.44 \\
0.295 & $^{226}\text{Ra}$ & 5.47 \\
0.336 & $^{232}\text{Th}$ & 6.31 \\
0.350 & $^{226}\text{Ra}$ & 8.33 \\
0.460 & Unidentified & 2.77 \\
1.173 & $^{60}\text{Co}$ & 11.85 \\
1.332 & $^{60}\text{Co}$ & 3.66 \\
1.461 & $^{40}\text{K}$ & 6.85 \\
2.614 & $^{208}\text{Tl}$ & 5.58 \\
 \hline
\end{tabular}
\caption{As a concrete example for the expected high-energy photon  background spectrum and magnitude, we consider the photon energies ($E_{\gamma_i}$) and densities ($n_{\gamma_i}$) and the radioactive contaminants based on the simulated EDELWEISS  background model which fits well with data collected in their detector setup at Modane~\cite{EDELWEISS:2018tde, Scorza:2015vla}. 
}
\label{tab:photondensities}
\end{table}

Assuming a flat Compton background created by photons of energy $E_{\gamma_i}$ with respective densities $n_{{\gamma}_i}$ given in  Table~\ref{tab:photondensities}, the total Compton rate at low energies in EDELWEISS' germanium target for an exposure of 1~kg-day is 
\begin{align}
\frac{dN_{\text{Comp}}}{dE_e} &= \sum_{i} \frac{1}{E_{\gamma_i}} \times \Big( \frac{1~\text{kg}}{M_{\text{Ge}}}   \Big) \times n_{\gamma_i} \sigma_{\text{Comp}}(E_{\gamma_i}) v_{\gamma} \times \text{day} \\ &\sim 0.042 ~~ \text{keV}^{-1}\,.
\end{align}
Here $\sigma_{\text{Comp}}$ is the Compton cross-section, and $M_{\text{Ge}}$ is the mass of a germanium atom. Our estimate of the rate of 0.042 events/kg/day/keV, calculated from the densities extracted by using the method described above, matches up to a factor of 2 with the flat Compton background at low energies reported by the EDELWEISS collaboration~\cite{EDELWEISS:2018tde}. 

\begin{figure}[!t] 
    \adjustimage{width=0.5\textwidth,center}{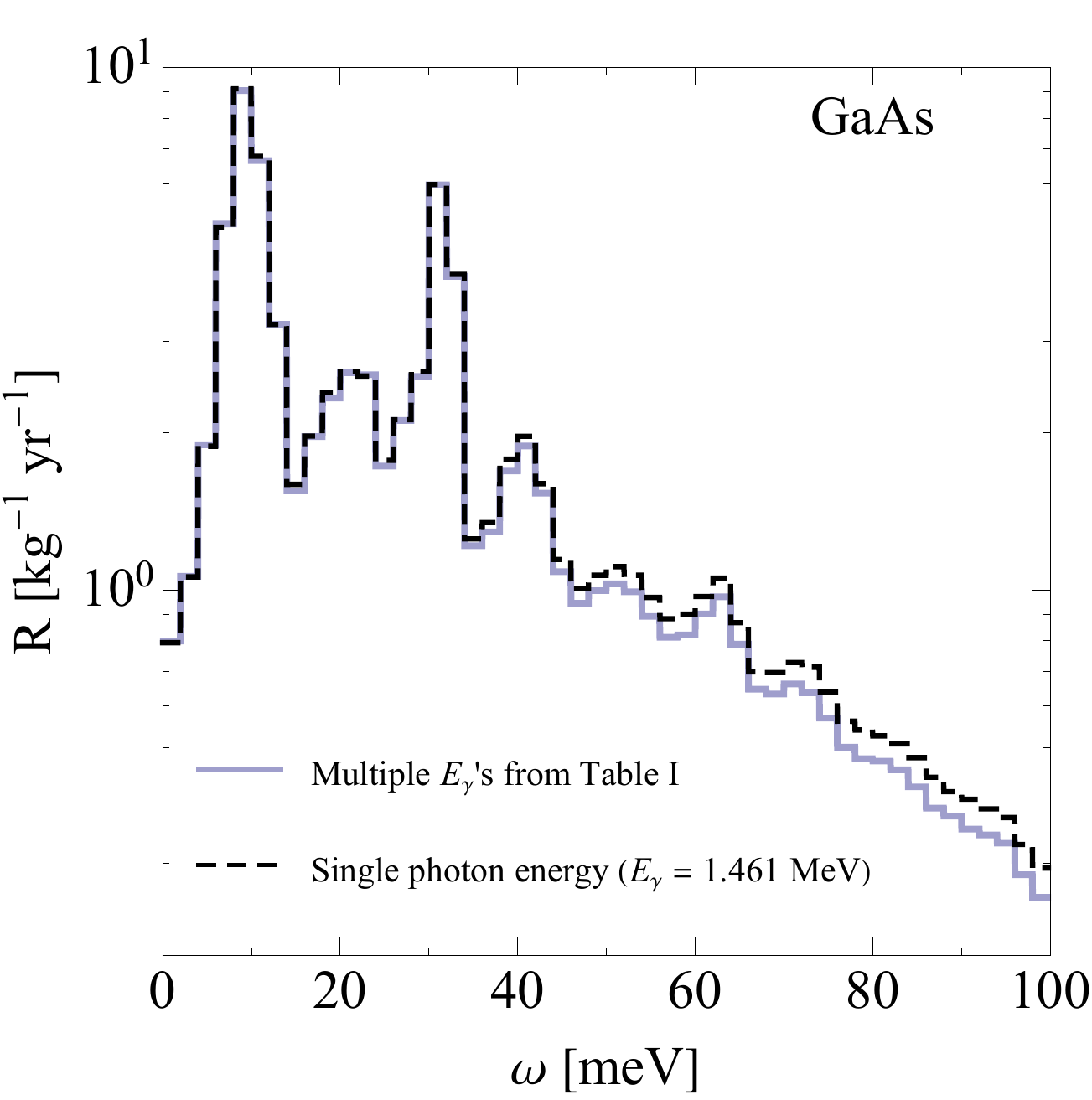}
    \vspace{-0.7cm}
    \caption{ Comparison of the phonon background spectrum in GaAs generated by the realistic combination of photon energies and densities given in Table~\ref{tab:photondensities} (pale blue line) versus with the spectrum generated by a single photon energy $E_{\gamma} = 1.461\, \text{MeV}$ with a photon density $n_{\gamma} = 1.3 \times 10^{-15} \, \text{cm}^{-1}$ chosen such that the normalization of the overall background agrees in the low-energy bins (black dashed line). 
    } \label{fig:comparison}
\end{figure} 

\begin{figure}[!t]
    \adjustimage{width=0.5\textwidth,center}{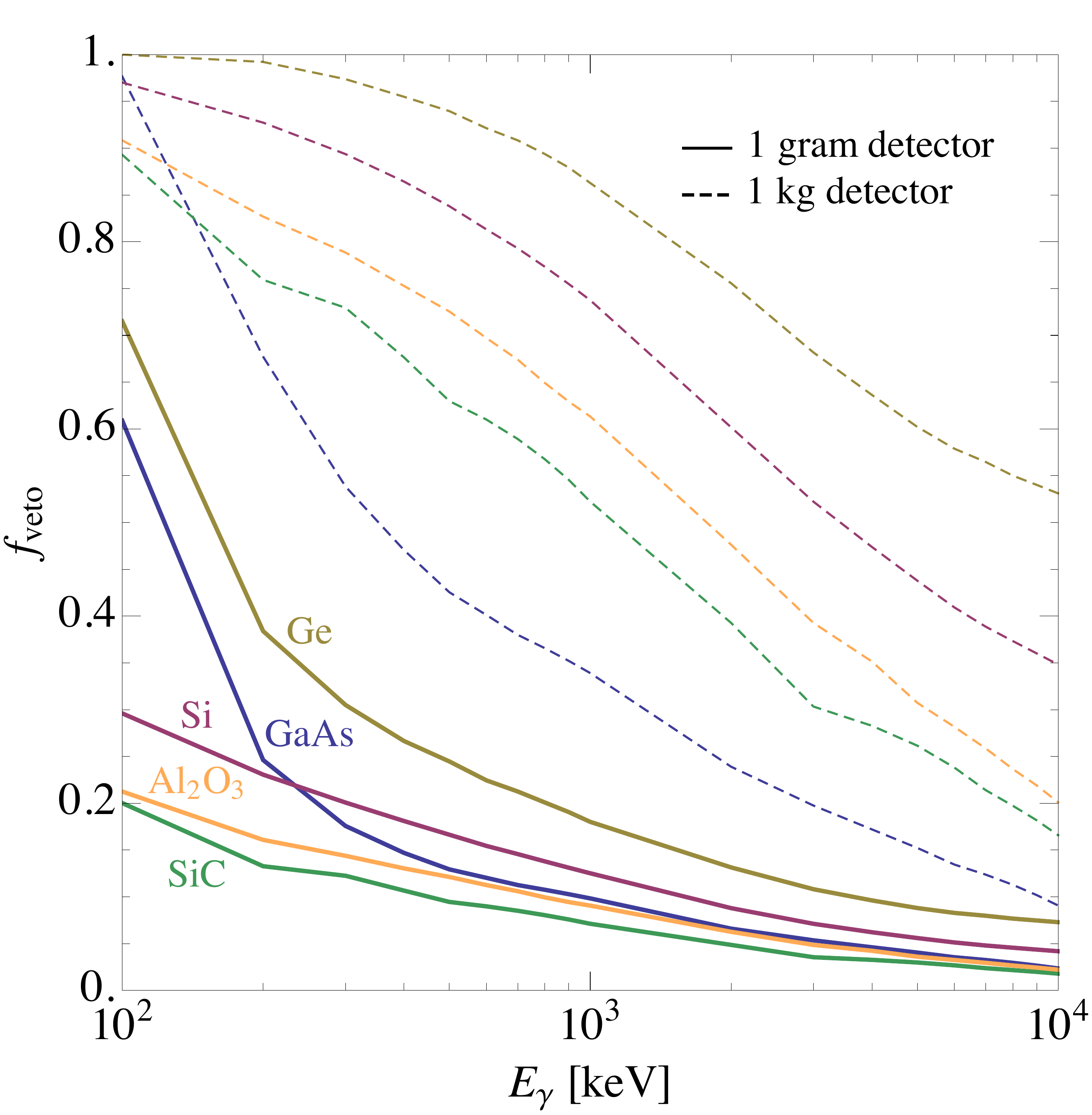}
    \vspace{-0.7cm}
    \caption{The fraction, $f_{\text{veto}}$, of phonon background events generated by high-energy photons that can be vetoed with the DM detector alone due to the high-energy photon Compton scattering or being absorbed by the DM detector.   We consider a cubic target material consisting of either GaAs, Ge, Si, SiC, and Al$_2$O$_3$, and a detector mass of 1~gram (solid lines) and 1~kg (dashed lines). 
     } 
     \label{fig:veto_fraction}
\end{figure} 

While we use the breakdown of photon energies and densities shown in Table~\ref{tab:photondensities} to show the main result in this work in Figs.~\ref{fig:background_rates} and \ref{fig:signal_rates}, we note that the spectrum itself (excluding the overall normalization) is insensitive to the assumed photon energies.  For example, we compare in Fig.~\ref{fig:comparison} the phonon spectrum generated by the realistic combination of photon energies and densities given in Table~\ref{tab:photondensities} versus with the spectrum generated by a single photon energy $E_{\gamma} = 1.461\, \text{MeV}$ with a photon density $n_{\gamma} = 1.3 \times10^{-15} \, \text{cm}^{-1}$ chosen such that the normalization of the overall background agrees in the low-energy bins.  We see that the spectra are very similar. If a detector design is able to achieve shielding beyond  $0.042$ count/kg/day/keV, the decrease of the expected background normalization depends on which photon energy densities are being reduced.  A linear reduction in the number densities shown in Table~\ref{tab:photondensities} for all energies would lead to a linear reduction in the total expected background. However, a reduction in lower energy photons is practically more feasible due to the larger total cross section, in which case one can estimate the relative decrease by noting that the contributions scale as $ \sim \sum_i  \frac{n_{E_{\gamma_i}}}{E^2_{\gamma_i}}$.

\section{Vetoing High-Energy Photons}\label{app:veto}
High energy photons can also interact in the detector through Compton scattering or photo-absorption.  If the photon is absorbed on its way into the target, then it does not create a phonon signal.  If a photon Compton scatters before or after the production of phonons, or is absorbed after creating a phonon signal, there will potentially be a high-energy event along with the low-energy phonon event. If the detector has timing information, then the low-energy event can be vetoed based on timing correlation with the high-energy event. Overall, these effects lead to a suppression of low-energy events that remain after the veto. This suppression depends on the mean free path of photons and the geometry of the detector and the presence of an active veto. 

If we do not consider the presence of a veto detector surrounding the DM detector, we can estimate the fraction of phonon background events that can be vetoed with the DM detector itself.  Let the mean free path of absorption of photons be $\lambda_\text{abs}$, and that of Compton scattering be $\lambda_\text{comp}$. Let the average length a photon has to travel inside the detector be $l_{\text{det}}$, and the fraction of events that do not happen due to prior photoabsorption or that can be vetoed be denoted by $f_{\text{veto}}$. Then $f_{\text{veto}}$ is given by 
\begin{equation}
f_{\text{veto}} \sim 1- e^{-l_{\text{det}}/\lambda_{\text{tot}}}\,,     
\end{equation}
where $\lambda_{\text{tot}}$ is the combined mean free path of producing a high energy event, given by
\begin{equation}
\lambda_{\text{tot}} = (\lambda_{\text{abs}}^{-1}+\lambda_{\text{comp}}^{-1})^{-1}\,.  
\end{equation}

In Fig.~\ref{fig:veto_fraction}, we show $f_{\text{veto}}$ for GaAs, Ge, Si, SiC, and Al$_2$O$_3$ for a 1~gram and a 1~kg detector assuming a cubic detector geometry.  We see that a large fraction of events cannot be vetoed by the DM detector alone, and that mitigating the coherent photon background will likely require the use of an active veto. 
 
\section{The Solar Neutrino Background}
\label{subsec:solar_neutrin} 
The dominant neutrino background for sub-GeV DM interactions arises from solar neutrinos via coherent neutrino-nucleus scattering~\cite{Essig:2018tss}. Various processes contribute to the total differential solar neutrino flux over the range of energy between $0.1 - 16$~MeV. The dominant flux component for the neutrino background shown in Fig.~\ref{fig:signal_rates} is due to the pp process ($p + p \to {}^2 {\rm H} + e^+ + \nu_e)$ with a total flux of $6 \times 10^{10} \, \text{cm}^{-1} \, \text{s}^{-1}$ over the energy range $0.1 - 0.43 \, \text{MeV}$. We take the differential fluxes provided $\frac{d\Phi_{\nu}}{dE_\nu}$ in~\cite{Bahcall:2004pz} to calculate the differential rate shown in Fig.~\ref{fig:signal_rates}, 
\begin{equation}
\frac{dR_{\nu}}{d\omega} (\omega) = N_T \int dE_{\nu} \int dq \frac{d\sigma_{N\nu}}{dqd\omega}(\vec{q},E_{\nu},\omega) \frac{d\Phi_{\nu}}{dE_\nu}\,,
\end{equation}
where the cross-section is
\begin{equation}
\frac{d\sigma_{N\nu}}{dqd\omega}(\vec{q},E_{\nu},\omega)  = \frac{d\sigma_{N\nu}}{dq}(\vec{q},E_{\nu})\, S(\vec{q},\omega)\,, 
\end{equation}
with 
\begin{equation}
\frac{d\sigma_{N\nu}}{dq} = \frac{G^2_F}{4\pi} q \left(N-Z\left(1-4 \sin \theta^2_w\right)\right)^2 \left(1-\frac{q^2}{4E^2_\nu}\right)F^2(q)  \,, 
\end{equation}
and $S(\vec{q},\omega)$ given by Eq.~\eqref{PDSF}. Here $G_F$ denotes the Fermi constant, $N$ is the number of neutrons per target nucleus, $Z$ is the number of protons per target nucleus, $\theta_w$ is the weak mixing angle, and $F(q)$ is the Helm form factor~\cite{LewinSmith}, which can be taken to be one at the momentum transfers relevant for phonon and multi-phonon signals.

 \bibliography{mybib}

\begin{thebibliography}{88}%
\makeatletter
\providecommand \@ifxundefined [1]{%
 \@ifx{#1\undefined}
}%
\providecommand \@ifnum [1]{%
 \ifnum #1\expandafter \@firstoftwo
 \else \expandafter \@secondoftwo
 \fi
}%
\providecommand \@ifx [1]{%
 \ifx #1\expandafter \@firstoftwo
 \else \expandafter \@secondoftwo
 \fi
}%
\providecommand \natexlab [1]{#1}%
\providecommand \enquote  [1]{``#1''}%
\providecommand \bibnamefont  [1]{#1}%
\providecommand \bibfnamefont [1]{#1}%
\providecommand \citenamefont [1]{#1}%
\providecommand \href@noop [0]{\@secondoftwo}%
\providecommand \href [0]{\begingroup \@sanitize@url \@href}%
\providecommand \@href[1]{\@@startlink{#1}\@@href}%
\providecommand \@@href[1]{\endgroup#1\@@endlink}%
\providecommand \@sanitize@url [0]{\catcode `\\12\catcode `\$12\catcode
  `\&12\catcode `\#12\catcode `\^12\catcode `\_12\catcode `\%12\relax}%
\providecommand \@@startlink[1]{}%
\providecommand \@@endlink[0]{}%
\providecommand \url  [0]{\begingroup\@sanitize@url \@url }%
\providecommand \@url [1]{\endgroup\@href {#1}{\urlprefix }}%
\providecommand \urlprefix  [0]{URL }%
\providecommand \Eprint [0]{\href }%
\providecommand \doibase [0]{https://doi.org/}%
\providecommand \selectlanguage [0]{\@gobble}%
\providecommand \bibinfo  [0]{\@secondoftwo}%
\providecommand \bibfield  [0]{\@secondoftwo}%
\providecommand \translation [1]{[#1]}%
\providecommand \BibitemOpen [0]{}%
\providecommand \bibitemStop [0]{}%
\providecommand \bibitemNoStop [0]{.\EOS\space}%
\providecommand \EOS [0]{\spacefactor3000\relax}%
\providecommand \BibitemShut  [1]{\csname bibitem#1\endcsname}%
\let\auto@bib@innerbib\@empty
\bibitem [{\citenamefont {Angle}\ \emph {et~al.}(2011)\citenamefont {Angle}
  \emph {et~al.}}]{Angle:2011th}%
  \BibitemOpen
  \bibfield  {author} {\bibinfo {author} {\bibfnamefont {J.}~\bibnamefont
  {Angle}} \emph {et~al.} (\bibinfo {collaboration} {XENON10}),\ }\bibfield
  {title} {\bibinfo {title} {{A Search for Light Dark Matter in Xenon10
  Data}},\ }\href {https://doi.org/10.1103/PhysRevLett.110.249901,
  10.1103/PhysRevLett.107.051301} {\bibfield  {journal} {\bibinfo  {journal}
  {Phys. Rev. Lett.}\ }\textbf {\bibinfo {volume} {107}},\ \bibinfo {pages}
  {051301} (\bibinfo {year} {2011})},\ \bibinfo {note} {[Erratum: Phys. Rev.
  Lett.110,249901(2013)]},\ \Eprint {https://arxiv.org/abs/1104.3088}
  {arXiv:1104.3088 [astro-ph.CO]} \BibitemShut {NoStop}%
\bibitem [{\citenamefont {Essig}\ \emph
  {et~al.}(2012{\natexlab{a}})\citenamefont {Essig}, \citenamefont
  {Manalaysay}, \citenamefont {Mardon}, \citenamefont {Sorensen},\ and\
  \citenamefont {Volansky}}]{Essig:2012yx}%
  \BibitemOpen
  \bibfield  {author} {\bibinfo {author} {\bibfnamefont {R.}~\bibnamefont
  {Essig}}, \bibinfo {author} {\bibfnamefont {A.}~\bibnamefont {Manalaysay}},
  \bibinfo {author} {\bibfnamefont {J.}~\bibnamefont {Mardon}}, \bibinfo
  {author} {\bibfnamefont {P.}~\bibnamefont {Sorensen}},\ and\ \bibinfo
  {author} {\bibfnamefont {T.}~\bibnamefont {Volansky}},\ }\bibfield  {title}
  {\bibinfo {title} {{First Direct Detection Limits on Sub-Gev Dark Matter from
  Xenon10}},\ }\href {https://doi.org/10.1103/PhysRevLett.109.021301}
  {\bibfield  {journal} {\bibinfo  {journal} {Phys. Rev. Lett.}\ }\textbf
  {\bibinfo {volume} {109}},\ \bibinfo {pages} {021301} (\bibinfo {year}
  {2012}{\natexlab{a}})},\ \Eprint {https://arxiv.org/abs/1206.2644}
  {arXiv:1206.2644 [astro-ph.CO]} \BibitemShut {NoStop}%
\bibitem [{\citenamefont {Hochberg}\ \emph
  {et~al.}(2016{\natexlab{a}})\citenamefont {Hochberg}, \citenamefont {Zhao},\
  and\ \citenamefont {Zurek}}]{Hochberg:2015pha}%
  \BibitemOpen
  \bibfield  {author} {\bibinfo {author} {\bibfnamefont {Y.}~\bibnamefont
  {Hochberg}}, \bibinfo {author} {\bibfnamefont {Y.}~\bibnamefont {Zhao}},\
  and\ \bibinfo {author} {\bibfnamefont {K.~M.}\ \bibnamefont {Zurek}},\
  }\bibfield  {title} {\bibinfo {title} {{Superconducting Detectors for
  Superlight Dark Matter}},\ }\href
  {https://doi.org/10.1103/PhysRevLett.116.011301} {\bibfield  {journal}
  {\bibinfo  {journal} {Phys. Rev. Lett.}\ }\textbf {\bibinfo {volume} {116}},\
  \bibinfo {pages} {011301} (\bibinfo {year} {2016}{\natexlab{a}})},\ \Eprint
  {https://arxiv.org/abs/1504.07237} {arXiv:1504.07237 [hep-ph]} \BibitemShut
  {NoStop}%
\bibitem [{\citenamefont {Hochberg}\ \emph
  {et~al.}(2016{\natexlab{b}})\citenamefont {Hochberg}, \citenamefont {Pyle},
  \citenamefont {Zhao},\ and\ \citenamefont {Zurek}}]{Hochberg:2015fth}%
  \BibitemOpen
  \bibfield  {author} {\bibinfo {author} {\bibfnamefont {Y.}~\bibnamefont
  {Hochberg}}, \bibinfo {author} {\bibfnamefont {M.}~\bibnamefont {Pyle}},
  \bibinfo {author} {\bibfnamefont {Y.}~\bibnamefont {Zhao}},\ and\ \bibinfo
  {author} {\bibfnamefont {K.~M.}\ \bibnamefont {Zurek}},\ }\bibfield  {title}
  {\bibinfo {title} {{Detecting Superlight Dark Matter with Fermi-Degenerate
  Materials}},\ }\href {https://doi.org/10.1007/JHEP08(2016)057} {\bibfield
  {journal} {\bibinfo  {journal} {JHEP}\ }\textbf {\bibinfo {volume} {08}},\
  \bibinfo {pages} {057}},\ \Eprint {https://arxiv.org/abs/1512.04533}
  {arXiv:1512.04533 [hep-ph]} \BibitemShut {NoStop}%
\bibitem [{\citenamefont {Aprile}\ \emph {et~al.}(2016)\citenamefont {Aprile}
  \emph {et~al.}}]{Aprile:2016wwo}%
  \BibitemOpen
  \bibfield  {author} {\bibinfo {author} {\bibfnamefont {E.}~\bibnamefont
  {Aprile}} \emph {et~al.} (\bibinfo {collaboration} {XENON}),\ }\bibfield
  {title} {\bibinfo {title} {{Low-Mass Dark Matter Search Using Ionization
  Signals in Xenon100}},\ }\href {https://doi.org/10.1103/PhysRevD.94.092001,
  10.1103/PhysRevD.95.059901} {\bibfield  {journal} {\bibinfo  {journal} {Phys.
  Rev.}\ }\textbf {\bibinfo {volume} {D94}},\ \bibinfo {pages} {092001}
  (\bibinfo {year} {2016})},\ \bibinfo {note} {[Erratum: Phys.
  Rev.D95,no.5,059901(2017)]},\ \Eprint {https://arxiv.org/abs/1605.06262}
  {arXiv:1605.06262 [astro-ph.CO]} \BibitemShut {NoStop}%
\bibitem [{\citenamefont {Schutz}\ and\ \citenamefont
  {Zurek}(2016)}]{Schutz:2016tid}%
  \BibitemOpen
  \bibfield  {author} {\bibinfo {author} {\bibfnamefont {K.}~\bibnamefont
  {Schutz}}\ and\ \bibinfo {author} {\bibfnamefont {K.~M.}\ \bibnamefont
  {Zurek}},\ }\bibfield  {title} {\bibinfo {title} {{Detectability of Light
  Dark Matter with Superfluid Helium}},\ }\href
  {https://doi.org/10.1103/PhysRevLett.117.121302} {\bibfield  {journal}
  {\bibinfo  {journal} {Phys. Rev. Lett.}\ }\textbf {\bibinfo {volume} {117}},\
  \bibinfo {pages} {121302} (\bibinfo {year} {2016})},\ \Eprint
  {https://arxiv.org/abs/1604.08206} {arXiv:1604.08206 [hep-ph]} \BibitemShut
  {NoStop}%
\bibitem [{\citenamefont {Essig}\ \emph {et~al.}(2017)\citenamefont {Essig},
  \citenamefont {Volansky},\ and\ \citenamefont {Yu}}]{Essig:2017kqs}%
  \BibitemOpen
  \bibfield  {author} {\bibinfo {author} {\bibfnamefont {R.}~\bibnamefont
  {Essig}}, \bibinfo {author} {\bibfnamefont {T.}~\bibnamefont {Volansky}},\
  and\ \bibinfo {author} {\bibfnamefont {T.-T.}\ \bibnamefont {Yu}},\
  }\bibfield  {title} {\bibinfo {title} {{New Constraints and Prospects for
  Sub-Gev Dark Matter Scattering Off Electrons in Xenon}},\ }\href
  {https://doi.org/10.1103/PhysRevD.96.043017} {\bibfield  {journal} {\bibinfo
  {journal} {Phys. Rev.}\ }\textbf {\bibinfo {volume} {D96}},\ \bibinfo {pages}
  {043017} (\bibinfo {year} {2017})},\ \Eprint
  {https://arxiv.org/abs/1703.00910} {arXiv:1703.00910 [hep-ph]} \BibitemShut
  {NoStop}%
\bibitem [{\citenamefont {Armengaud}\ \emph {et~al.}(2017)\citenamefont
  {Armengaud} \emph {et~al.}}]{EDELWEISS:2017lvq}%
  \BibitemOpen
  \bibfield  {author} {\bibinfo {author} {\bibfnamefont {E.}~\bibnamefont
  {Armengaud}} \emph {et~al.} (\bibinfo {collaboration} {EDELWEISS}),\
  }\bibfield  {title} {\bibinfo {title} {{Performance of the EDELWEISS-III
  experiment for direct dark matter searches}},\ }\href
  {https://doi.org/10.1088/1748-0221/12/08/P08010} {\bibfield  {journal}
  {\bibinfo  {journal} {JINST}\ }\textbf {\bibinfo {volume} {12}}\bibfield
  {number} {\bibinfo  {number} { (08)},\ \bibinfo {pages} {P08010}},\ }\Eprint
  {https://arxiv.org/abs/1706.01070} {arXiv:1706.01070 [physics.ins-det]}
  \BibitemShut {NoStop}%
\bibitem [{\citenamefont {Tiffenberg}\ \emph {et~al.}(2017)\citenamefont
  {Tiffenberg}, \citenamefont {Sofo-Haro}, \citenamefont {Drlica-Wagner},
  \citenamefont {Essig}, \citenamefont {Guardincerri}, \citenamefont {Holland},
  \citenamefont {Volansky},\ and\ \citenamefont {Yu}}]{Tiffenberg:2017aac}%
  \BibitemOpen
  \bibfield  {author} {\bibinfo {author} {\bibfnamefont {J.}~\bibnamefont
  {Tiffenberg}}, \bibinfo {author} {\bibfnamefont {M.}~\bibnamefont
  {Sofo-Haro}}, \bibinfo {author} {\bibfnamefont {A.}~\bibnamefont
  {Drlica-Wagner}}, \bibinfo {author} {\bibfnamefont {R.}~\bibnamefont
  {Essig}}, \bibinfo {author} {\bibfnamefont {Y.}~\bibnamefont {Guardincerri}},
  \bibinfo {author} {\bibfnamefont {S.}~\bibnamefont {Holland}}, \bibinfo
  {author} {\bibfnamefont {T.}~\bibnamefont {Volansky}},\ and\ \bibinfo
  {author} {\bibfnamefont {T.-T.}\ \bibnamefont {Yu}} (\bibinfo {collaboration}
  {SENSEI}),\ }\bibfield  {title} {\bibinfo {title} {{Single-electron and
  single-photon sensitivity with a silicon Skipper CCD}},\ }\href
  {https://doi.org/10.1103/PhysRevLett.119.131802} {\bibfield  {journal}
  {\bibinfo  {journal} {Phys. Rev. Lett.}\ }\textbf {\bibinfo {volume} {119}},\
  \bibinfo {pages} {131802} (\bibinfo {year} {2017})},\ \Eprint
  {https://arxiv.org/abs/1706.00028} {arXiv:1706.00028 [physics.ins-det]}
  \BibitemShut {NoStop}%
\bibitem [{\citenamefont {Angloher}\ \emph {et~al.}(2017)\citenamefont
  {Angloher} \emph {et~al.}}]{Angloher:2017sxg}%
  \BibitemOpen
  \bibfield  {author} {\bibinfo {author} {\bibfnamefont {G.}~\bibnamefont
  {Angloher}} \emph {et~al.} (\bibinfo {collaboration} {CRESST}),\ }\bibfield
  {title} {\bibinfo {title} {{Results on MeV-scale dark matter from a
  gram-scale cryogenic calorimeter operated above ground}},\ }\href
  {https://doi.org/10.1140/epjc/s10052-017-5223-9} {\bibfield  {journal}
  {\bibinfo  {journal} {Eur. Phys. J. C}\ }\textbf {\bibinfo {volume} {77}},\
  \bibinfo {pages} {637} (\bibinfo {year} {2017})},\ \Eprint
  {https://arxiv.org/abs/1707.06749} {arXiv:1707.06749 [astro-ph.CO]}
  \BibitemShut {NoStop}%
\bibitem [{\citenamefont {Romani}\ \emph {et~al.}(2018)\citenamefont {Romani}
  \emph {et~al.}}]{Romani:2017iwi}%
  \BibitemOpen
  \bibfield  {author} {\bibinfo {author} {\bibfnamefont {R.~K.}\ \bibnamefont
  {Romani}} \emph {et~al.},\ }\bibfield  {title} {\bibinfo {title} {{Thermal
  Detection of Single E-H Pairs in a Biased Silicon Crystal Detector}},\ }\href
  {https://doi.org/10.1063/1.5010699} {\bibfield  {journal} {\bibinfo
  {journal} {Appl. Phys. Lett.}\ }\textbf {\bibinfo {volume} {112}},\ \bibinfo
  {pages} {043501} (\bibinfo {year} {2018})},\ \Eprint
  {https://arxiv.org/abs/1710.09335} {arXiv:1710.09335 [physics.ins-det]}
  \BibitemShut {NoStop}%
\bibitem [{\citenamefont {Hochberg}\ \emph {et~al.}(2018)\citenamefont
  {Hochberg}, \citenamefont {Kahn}, \citenamefont {Lisanti}, \citenamefont
  {Zurek}, \citenamefont {Grushin}, \citenamefont {Ilan}, \citenamefont
  {Griffin}, \citenamefont {Liu}, \citenamefont {Weber},\ and\ \citenamefont
  {Neaton}}]{Hochberg:2017wce}%
  \BibitemOpen
  \bibfield  {author} {\bibinfo {author} {\bibfnamefont {Y.}~\bibnamefont
  {Hochberg}}, \bibinfo {author} {\bibfnamefont {Y.}~\bibnamefont {Kahn}},
  \bibinfo {author} {\bibfnamefont {M.}~\bibnamefont {Lisanti}}, \bibinfo
  {author} {\bibfnamefont {K.~M.}\ \bibnamefont {Zurek}}, \bibinfo {author}
  {\bibfnamefont {A.~G.}\ \bibnamefont {Grushin}}, \bibinfo {author}
  {\bibfnamefont {R.}~\bibnamefont {Ilan}}, \bibinfo {author} {\bibfnamefont
  {S.~M.}\ \bibnamefont {Griffin}}, \bibinfo {author} {\bibfnamefont {Z.-F.}\
  \bibnamefont {Liu}}, \bibinfo {author} {\bibfnamefont {S.~F.}\ \bibnamefont
  {Weber}},\ and\ \bibinfo {author} {\bibfnamefont {J.~B.}\ \bibnamefont
  {Neaton}},\ }\bibfield  {title} {\bibinfo {title} {{Detection of sub-MeV Dark
  Matter with Three-Dimensional Dirac Materials}},\ }\href
  {https://doi.org/10.1103/PhysRevD.97.015004} {\bibfield  {journal} {\bibinfo
  {journal} {Phys. Rev. D}\ }\textbf {\bibinfo {volume} {97}},\ \bibinfo
  {pages} {015004} (\bibinfo {year} {2018})},\ \Eprint
  {https://arxiv.org/abs/1708.08929} {arXiv:1708.08929 [hep-ph]} \BibitemShut
  {NoStop}%
\bibitem [{\citenamefont {Cavoto}\ \emph {et~al.}(2018)\citenamefont {Cavoto},
  \citenamefont {Luchetta},\ and\ \citenamefont {Polosa}}]{Cavoto:2017otc}%
  \BibitemOpen
  \bibfield  {author} {\bibinfo {author} {\bibfnamefont {G.}~\bibnamefont
  {Cavoto}}, \bibinfo {author} {\bibfnamefont {F.}~\bibnamefont {Luchetta}},\
  and\ \bibinfo {author} {\bibfnamefont {A.~D.}\ \bibnamefont {Polosa}},\
  }\bibfield  {title} {\bibinfo {title} {{Sub-GeV Dark Matter Detection with
  Electron Recoils in Carbon Nanotubes}},\ }\href
  {https://doi.org/10.1016/j.physletb.2017.11.064} {\bibfield  {journal}
  {\bibinfo  {journal} {Phys. Lett. B}\ }\textbf {\bibinfo {volume} {776}},\
  \bibinfo {pages} {338} (\bibinfo {year} {2018})},\ \Eprint
  {https://arxiv.org/abs/1706.02487} {arXiv:1706.02487 [hep-ph]} \BibitemShut
  {NoStop}%
\bibitem [{\citenamefont {Agnes}\ \emph {et~al.}(2018)\citenamefont {Agnes}
  \emph {et~al.}}]{Agnes:2018oej}%
  \BibitemOpen
  \bibfield  {author} {\bibinfo {author} {\bibfnamefont {P.}~\bibnamefont
  {Agnes}} \emph {et~al.} (\bibinfo {collaboration} {DarkSide}),\ }\bibfield
  {title} {\bibinfo {title} {{Constraints on Sub-GeV Dark-Matter-Electron
  Scattering from the DarkSide-50 Experiment}},\ }\href
  {https://doi.org/10.1103/PhysRevLett.121.111303} {\bibfield  {journal}
  {\bibinfo  {journal} {Phys. Rev. Lett.}\ }\textbf {\bibinfo {volume} {121}},\
  \bibinfo {pages} {111303} (\bibinfo {year} {2018})},\ \Eprint
  {https://arxiv.org/abs/1802.06998} {arXiv:1802.06998 [astro-ph.CO]}
  \BibitemShut {NoStop}%
\bibitem [{\citenamefont {Agnese}\ \emph
  {et~al.}(2018{\natexlab{a}})\citenamefont {Agnese} \emph
  {et~al.}}]{CDMS:2018nhx}%
  \BibitemOpen
  \bibfield  {author} {\bibinfo {author} {\bibfnamefont {R.}~\bibnamefont
  {Agnese}} \emph {et~al.} (\bibinfo {collaboration} {CDMS}),\ }\bibfield
  {title} {\bibinfo {title} {{Nuclear-Recoil Energy Scale in CDMS II Silicon
  Dark-Matter Detectors}},\ }\href {https://doi.org/10.1016/j.nima.2018.07.028}
  {\bibfield  {journal} {\bibinfo  {journal} {Nucl. Instrum. Meth. A}\ }\textbf
  {\bibinfo {volume} {905}},\ \bibinfo {pages} {71} (\bibinfo {year}
  {2018}{\natexlab{a}})},\ \Eprint {https://arxiv.org/abs/1803.02903}
  {arXiv:1803.02903 [physics.ins-det]} \BibitemShut {NoStop}%
\bibitem [{\citenamefont {Agnese}\ \emph
  {et~al.}(2018{\natexlab{b}})\citenamefont {Agnese} \emph
  {et~al.}}]{Agnese:2018col}%
  \BibitemOpen
  \bibfield  {author} {\bibinfo {author} {\bibfnamefont {R.}~\bibnamefont
  {Agnese}} \emph {et~al.} (\bibinfo {collaboration} {SuperCDMS}),\ }\bibfield
  {title} {\bibinfo {title} {{First Dark Matter Constraints from a Supercdms
  Single-Charge Sensitive Detector}},\ }\href
  {https://doi.org/10.1103/PhysRevLett.121.051301} {\bibfield  {journal}
  {\bibinfo  {journal} {Phys. Rev. Lett.}\ }\textbf {\bibinfo {volume} {121}},\
  \bibinfo {pages} {051301} (\bibinfo {year} {2018}{\natexlab{b}})},\ \Eprint
  {https://arxiv.org/abs/1804.10697} {arXiv:1804.10697 [hep-ex]} \BibitemShut
  {NoStop}%
\bibitem [{\citenamefont {Aguilar-Arevalo}\ \emph {et~al.}(2019)\citenamefont
  {Aguilar-Arevalo} \emph {et~al.}}]{Aguilar-Arevalo:2019wdi}%
  \BibitemOpen
  \bibfield  {author} {\bibinfo {author} {\bibfnamefont {A.}~\bibnamefont
  {Aguilar-Arevalo}} \emph {et~al.} (\bibinfo {collaboration} {DAMIC}),\
  }\bibfield  {title} {\bibinfo {title} {{Constraints on Light Dark Matter
  Particles Interacting with Electrons from Damic at Snolab}},\ }\href
  {https://doi.org/10.1103/PhysRevLett.123.181802} {\bibfield  {journal}
  {\bibinfo  {journal} {Phys. Rev. Lett.}\ }\textbf {\bibinfo {volume} {123}},\
  \bibinfo {pages} {181802} (\bibinfo {year} {2019})},\ \Eprint
  {https://arxiv.org/abs/1907.12628} {arXiv:1907.12628 [astro-ph.CO]}
  \BibitemShut {NoStop}%
\bibitem [{\citenamefont {Essig}\ \emph {et~al.}(2019)\citenamefont {Essig},
  \citenamefont {P\'erez-R\'\i{}os}, \citenamefont {Ramani},\ and\
  \citenamefont {Slone}}]{Essig:2019kfe}%
  \BibitemOpen
  \bibfield  {author} {\bibinfo {author} {\bibfnamefont {R.}~\bibnamefont
  {Essig}}, \bibinfo {author} {\bibfnamefont {J.}~\bibnamefont
  {P\'erez-R\'\i{}os}}, \bibinfo {author} {\bibfnamefont {H.}~\bibnamefont
  {Ramani}},\ and\ \bibinfo {author} {\bibfnamefont {O.}~\bibnamefont
  {Slone}},\ }\bibfield  {title} {\bibinfo {title} {{Direct Detection of
  Spin-(In)dependent Nuclear Scattering of Sub-GeV Dark Matter Using Molecular
  Excitations}},\ }\href {https://doi.org/10.1103/PhysRevResearch.1.033105}
  {\bibfield  {journal} {\bibinfo  {journal} {Phys. Rev. Research.}\ }\textbf
  {\bibinfo {volume} {1}},\ \bibinfo {pages} {033105} (\bibinfo {year}
  {2019})},\ \Eprint {https://arxiv.org/abs/1907.07682} {arXiv:1907.07682
  [hep-ph]} \BibitemShut {NoStop}%
\bibitem [{\citenamefont {Barak}\ \emph
  {et~al.}(2020{\natexlab{a}})\citenamefont {Barak}, \citenamefont {Bloch},
  \citenamefont {Cababie}, \citenamefont {Cancelo}, \citenamefont {Chaplinsky},
  \citenamefont {Chierchie}, \citenamefont {Crisler}, \citenamefont
  {Drlica-Wagner}, \citenamefont {Essig}, \citenamefont {Estrada},
  \citenamefont {Etzion}, \citenamefont {Moroni}, \citenamefont {Gift},
  \citenamefont {Munagavalasa}, \citenamefont {Orly}, \citenamefont
  {Rodrigues}, \citenamefont {Singal}, \citenamefont {Haro}, \citenamefont
  {Stefanazzi}, \citenamefont {Tiffenberg}, \citenamefont {Uemura},
  \citenamefont {Volansky},\ and\ \citenamefont {Yu}}]{Physicality.125.171802}%
  \BibitemOpen
  \bibfield  {author} {\bibinfo {author} {\bibfnamefont {L.}~\bibnamefont
  {Barak}}, \bibinfo {author} {\bibfnamefont {I.~M.}\ \bibnamefont {Bloch}},
  \bibinfo {author} {\bibfnamefont {M.}~\bibnamefont {Cababie}}, \bibinfo
  {author} {\bibfnamefont {G.}~\bibnamefont {Cancelo}}, \bibinfo {author}
  {\bibfnamefont {L.}~\bibnamefont {Chaplinsky}}, \bibinfo {author}
  {\bibfnamefont {F.}~\bibnamefont {Chierchie}}, \bibinfo {author}
  {\bibfnamefont {M.}~\bibnamefont {Crisler}}, \bibinfo {author} {\bibfnamefont
  {A.}~\bibnamefont {Drlica-Wagner}}, \bibinfo {author} {\bibfnamefont
  {R.}~\bibnamefont {Essig}}, \bibinfo {author} {\bibfnamefont
  {J.}~\bibnamefont {Estrada}}, \bibinfo {author} {\bibfnamefont
  {E.}~\bibnamefont {Etzion}}, \bibinfo {author} {\bibfnamefont {G.~F.}\
  \bibnamefont {Moroni}}, \bibinfo {author} {\bibfnamefont {D.}~\bibnamefont
  {Gift}}, \bibinfo {author} {\bibfnamefont {S.}~\bibnamefont {Munagavalasa}},
  \bibinfo {author} {\bibfnamefont {A.}~\bibnamefont {Orly}}, \bibinfo {author}
  {\bibfnamefont {D.}~\bibnamefont {Rodrigues}}, \bibinfo {author}
  {\bibfnamefont {A.}~\bibnamefont {Singal}}, \bibinfo {author} {\bibfnamefont
  {M.~S.}\ \bibnamefont {Haro}}, \bibinfo {author} {\bibfnamefont
  {L.}~\bibnamefont {Stefanazzi}}, \bibinfo {author} {\bibfnamefont
  {J.}~\bibnamefont {Tiffenberg}}, \bibinfo {author} {\bibfnamefont
  {S.}~\bibnamefont {Uemura}}, \bibinfo {author} {\bibfnamefont
  {T.}~\bibnamefont {Volansky}},\ and\ \bibinfo {author} {\bibfnamefont
  {T.-T.}\ \bibnamefont {Yu}} (\bibinfo {collaboration} {SENSEI
  Collaboration}),\ }\bibfield  {title} {\bibinfo {title} {Sensei:
  Direct-detection results on sub-gev dark matter from a new skipper ccd},\
  }\href {https://doi.org/10.1103/PhysRevLett.125.171802} {\bibfield  {journal}
  {\bibinfo  {journal} {Phys. Rev. Lett.}\ }\textbf {\bibinfo {volume} {125}},\
  \bibinfo {pages} {171802} (\bibinfo {year} {2020}{\natexlab{a}})}\BibitemShut
  {NoStop}%
\bibitem [{\citenamefont {Aprile}\ \emph
  {et~al.}(2019{\natexlab{a}})\citenamefont {Aprile} \emph
  {et~al.}}]{Aprile:2019xxb}%
  \BibitemOpen
  \bibfield  {author} {\bibinfo {author} {\bibfnamefont {E.}~\bibnamefont
  {Aprile}} \emph {et~al.} (\bibinfo {collaboration} {XENON}),\ }\bibfield
  {title} {\bibinfo {title} {{Light Dark Matter Search with Ionization Signals
  in Xenon1T}},\ }\href {https://doi.org/10.1103/PhysRevLett.123.251801}
  {\bibfield  {journal} {\bibinfo  {journal} {Phys. Rev. Lett.}\ }\textbf
  {\bibinfo {volume} {123}},\ \bibinfo {pages} {251801} (\bibinfo {year}
  {2019}{\natexlab{a}})},\ \Eprint {https://arxiv.org/abs/1907.11485}
  {arXiv:1907.11485 [hep-ex]} \BibitemShut {NoStop}%
\bibitem [{\citenamefont {Amaral}\ \emph {et~al.}(2020)\citenamefont {Amaral}
  \emph {et~al.}}]{Amaral:2020ryn}%
  \BibitemOpen
  \bibfield  {author} {\bibinfo {author} {\bibfnamefont {D.~W.}\ \bibnamefont
  {Amaral}} \emph {et~al.} (\bibinfo {collaboration} {SuperCDMS}),\ }\bibfield
  {title} {\bibinfo {title} {{Constraints on low-mass, relic dark matter
  candidates from a surface-operated SuperCDMS single-charge sensitive
  detector}},\ }\href {https://doi.org/10.1103/PhysRevD.102.091101} {\bibfield
  {journal} {\bibinfo  {journal} {Phys. Rev. D}\ }\textbf {\bibinfo {volume}
  {102}},\ \bibinfo {pages} {091101} (\bibinfo {year} {2020})},\ \Eprint
  {https://arxiv.org/abs/2005.14067} {arXiv:2005.14067 [hep-ex]} \BibitemShut
  {NoStop}%
\bibitem [{\citenamefont {Aprile}\ \emph
  {et~al.}(2019{\natexlab{b}})\citenamefont {Aprile} \emph
  {et~al.}}]{XENON:2019gfn}%
  \BibitemOpen
  \bibfield  {author} {\bibinfo {author} {\bibfnamefont {E.}~\bibnamefont
  {Aprile}} \emph {et~al.} (\bibinfo {collaboration} {XENON}),\ }\bibfield
  {title} {\bibinfo {title} {{Light Dark Matter Search with Ionization Signals
  in XENON1T}},\ }\href {https://doi.org/10.1103/PhysRevLett.123.251801}
  {\bibfield  {journal} {\bibinfo  {journal} {Phys. Rev. Lett.}\ }\textbf
  {\bibinfo {volume} {123}},\ \bibinfo {pages} {251801} (\bibinfo {year}
  {2019}{\natexlab{b}})},\ \Eprint {https://arxiv.org/abs/1907.11485}
  {arXiv:1907.11485 [hep-ex]} \BibitemShut {NoStop}%
\bibitem [{\citenamefont {Hochberg}\ \emph {et~al.}(2019)\citenamefont
  {Hochberg}, \citenamefont {Charaev}, \citenamefont {Nam}, \citenamefont
  {Verma}, \citenamefont {Colangelo},\ and\ \citenamefont
  {Berggren}}]{Hochberg:2019cyy}%
  \BibitemOpen
  \bibfield  {author} {\bibinfo {author} {\bibfnamefont {Y.}~\bibnamefont
  {Hochberg}}, \bibinfo {author} {\bibfnamefont {I.}~\bibnamefont {Charaev}},
  \bibinfo {author} {\bibfnamefont {S.-W.}\ \bibnamefont {Nam}}, \bibinfo
  {author} {\bibfnamefont {V.}~\bibnamefont {Verma}}, \bibinfo {author}
  {\bibfnamefont {M.}~\bibnamefont {Colangelo}},\ and\ \bibinfo {author}
  {\bibfnamefont {K.~K.}\ \bibnamefont {Berggren}},\ }\bibfield  {title}
  {\bibinfo {title} {{Detecting Sub-GeV Dark Matter with Superconducting
  Nanowires}},\ }\href {https://doi.org/10.1103/PhysRevLett.123.151802}
  {\bibfield  {journal} {\bibinfo  {journal} {Phys. Rev. Lett.}\ }\textbf
  {\bibinfo {volume} {123}},\ \bibinfo {pages} {151802} (\bibinfo {year}
  {2019})},\ \Eprint {https://arxiv.org/abs/1903.05101} {arXiv:1903.05101
  [hep-ph]} \BibitemShut {NoStop}%
\bibitem [{\citenamefont {Abdelhameed}\ \emph
  {et~al.}(2019{\natexlab{a}})\citenamefont {Abdelhameed} \emph
  {et~al.}}]{Abdelhameed:2019hmk}%
  \BibitemOpen
  \bibfield  {author} {\bibinfo {author} {\bibfnamefont {A.}~\bibnamefont
  {Abdelhameed}} \emph {et~al.} (\bibinfo {collaboration} {CRESST}),\
  }\bibfield  {title} {\bibinfo {title} {{First results from the CRESST-III
  low-mass dark matter program}},\ }\href
  {https://doi.org/10.1103/PhysRevD.100.102002} {\bibfield  {journal} {\bibinfo
   {journal} {Phys. Rev. D}\ }\textbf {\bibinfo {volume} {100}},\ \bibinfo
  {pages} {102002} (\bibinfo {year} {2019}{\natexlab{a}})},\ \Eprint
  {https://arxiv.org/abs/1904.00498} {arXiv:1904.00498 [astro-ph.CO]}
  \BibitemShut {NoStop}%
\bibitem [{\citenamefont {Abdelhameed}\ \emph
  {et~al.}(2019{\natexlab{b}})\citenamefont {Abdelhameed} \emph
  {et~al.}}]{Abdelhameed:2019mac}%
  \BibitemOpen
  \bibfield  {author} {\bibinfo {author} {\bibfnamefont {A.~H.}\ \bibnamefont
  {Abdelhameed}} \emph {et~al.} (\bibinfo {collaboration} {CRESST}),\
  }\href@noop {} {\bibinfo {title} {{Description of CRESST-III Data}}},\
  \bibinfo {howpublished} {arXiv:1905.07335 [astro-ph.CO]} (\bibinfo {year}
  {2019}{\natexlab{b}}),\ \Eprint {https://arxiv.org/abs/1905.07335}
  {arXiv:1905.07335 [astro-ph.CO]} \BibitemShut {NoStop}%
\bibitem [{\citenamefont {Blanco}\ \emph {et~al.}(2020)\citenamefont {Blanco},
  \citenamefont {Collar}, \citenamefont {Kahn},\ and\ \citenamefont
  {Lillard}}]{Blanco:2019lrf}%
  \BibitemOpen
  \bibfield  {author} {\bibinfo {author} {\bibfnamefont {C.}~\bibnamefont
  {Blanco}}, \bibinfo {author} {\bibfnamefont {J.~I.}\ \bibnamefont {Collar}},
  \bibinfo {author} {\bibfnamefont {Y.}~\bibnamefont {Kahn}},\ and\ \bibinfo
  {author} {\bibfnamefont {B.}~\bibnamefont {Lillard}},\ }\bibfield  {title}
  {\bibinfo {title} {{Dark Matter-Electron Scattering from Aromatic Organic
  Targets}},\ }\href {https://doi.org/10.1103/PhysRevD.101.056001} {\bibfield
  {journal} {\bibinfo  {journal} {Phys. Rev. D}\ }\textbf {\bibinfo {volume}
  {101}},\ \bibinfo {pages} {056001} (\bibinfo {year} {2020})},\ \Eprint
  {https://arxiv.org/abs/1912.02822} {arXiv:1912.02822 [hep-ph]} \BibitemShut
  {NoStop}%
\bibitem [{\citenamefont {Barak}\ \emph
  {et~al.}(2020{\natexlab{b}})\citenamefont {Barak} \emph
  {et~al.}}]{SENSEI:2020dpa}%
  \BibitemOpen
  \bibfield  {author} {\bibinfo {author} {\bibfnamefont {L.}~\bibnamefont
  {Barak}} \emph {et~al.} (\bibinfo {collaboration} {SENSEI}),\ }\bibfield
  {title} {\bibinfo {title} {{SENSEI: Direct-Detection Results on sub-GeV Dark
  Matter from a New Skipper-CCD}},\ }\href
  {https://doi.org/10.1103/PhysRevLett.125.171802} {\bibfield  {journal}
  {\bibinfo  {journal} {Phys. Rev. Lett.}\ }\textbf {\bibinfo {volume} {125}},\
  \bibinfo {pages} {171802} (\bibinfo {year} {2020}{\natexlab{b}})},\ \Eprint
  {https://arxiv.org/abs/2004.11378} {arXiv:2004.11378 [astro-ph.CO]}
  \BibitemShut {NoStop}%
\bibitem [{\citenamefont {{D.\ S.\ Akerib {\it et al.} [LUX
  Collaboration]}}(2020)}]{LUX:2020vbj}%
  \BibitemOpen
  \bibfield  {author} {\bibinfo {author} {\bibnamefont {{D.\ S.\ Akerib {\it et
  al.} [LUX Collaboration]}}} (\bibinfo {collaboration} {LUX}),\ }\bibfield
  {title} {\bibinfo {title} {{Investigation of background electron emission in
  the LUX detector}},\ }\href {https://doi.org/10.1103/PhysRevD.102.092004}
  {\bibfield  {journal} {\bibinfo  {journal} {Phys. Rev. D}\ }\textbf {\bibinfo
  {volume} {102}},\ \bibinfo {pages} {092004} (\bibinfo {year} {2020})},\
  \Eprint {https://arxiv.org/abs/2004.07791} {arXiv:2004.07791} \BibitemShut
  {NoStop}%
\bibitem [{\citenamefont {Arnaud}\ \emph {et~al.}(2020)\citenamefont {Arnaud}
  \emph {et~al.}}]{Arnaud:2020svb}%
  \BibitemOpen
  \bibfield  {author} {\bibinfo {author} {\bibfnamefont {Q.}~\bibnamefont
  {Arnaud}} \emph {et~al.} (\bibinfo {collaboration} {EDELWEISS}),\ }\bibfield
  {title} {\bibinfo {title} {{First germanium-based constraints on sub-MeV Dark
  Matter with the EDELWEISS experiment}},\ }\href
  {https://doi.org/10.1103/PhysRevLett.125.141301} {\bibfield  {journal}
  {\bibinfo  {journal} {Phys. Rev. Lett.}\ }\textbf {\bibinfo {volume} {125}},\
  \bibinfo {pages} {141301} (\bibinfo {year} {2020})},\ \Eprint
  {https://arxiv.org/abs/2003.01046} {arXiv:2003.01046 [astro-ph.GA]}
  \BibitemShut {NoStop}%
\bibitem [{\citenamefont {Canonica}\ \emph {et~al.}(2020)\citenamefont
  {Canonica} \emph {et~al.}}]{Canonica:2020omq}%
  \BibitemOpen
  \bibfield  {author} {\bibinfo {author} {\bibfnamefont {L.}~\bibnamefont
  {Canonica}} \emph {et~al.},\ }\bibfield  {title} {\bibinfo {title}
  {{Operation of a Diamond Cryogenic Detector for Low-Mass Dark Matter
  Searches}},\ }\href {https://doi.org/10.1007/s10909-020-02350-4} {\bibfield
  {journal} {\bibinfo  {journal} {J. Low Temp. Phys.}\ }\textbf {\bibinfo
  {volume} {199}},\ \bibinfo {pages} {606} (\bibinfo {year}
  {2020})}\BibitemShut {NoStop}%
\bibitem [{\citenamefont {Alkhatib}\ \emph {et~al.}(2021)\citenamefont
  {Alkhatib} \emph {et~al.}}]{SuperCDMS:2020aus}%
  \BibitemOpen
  \bibfield  {author} {\bibinfo {author} {\bibfnamefont {I.}~\bibnamefont
  {Alkhatib}} \emph {et~al.} (\bibinfo {collaboration} {SuperCDMS}),\
  }\bibfield  {title} {\bibinfo {title} {{Light Dark Matter Search with a
  High-Resolution Athermal Phonon Detector Operated Above Ground}},\ }\href
  {https://doi.org/10.1103/PhysRevLett.127.061801} {\bibfield  {journal}
  {\bibinfo  {journal} {Phys. Rev. Lett.}\ }\textbf {\bibinfo {volume} {127}},\
  \bibinfo {pages} {061801} (\bibinfo {year} {2021})},\ \Eprint
  {https://arxiv.org/abs/2007.14289} {arXiv:2007.14289 [hep-ex]} \BibitemShut
  {NoStop}%
\bibitem [{\citenamefont {Hochberg}\ \emph
  {et~al.}(2021{\natexlab{a}})\citenamefont {Hochberg}, \citenamefont
  {Lehmann}, \citenamefont {Charaev}, \citenamefont {Chiles}, \citenamefont
  {Colangelo}, \citenamefont {Nam},\ and\ \citenamefont
  {Berggren}}]{Hochberg:2021yud}%
  \BibitemOpen
  \bibfield  {author} {\bibinfo {author} {\bibfnamefont {Y.}~\bibnamefont
  {Hochberg}}, \bibinfo {author} {\bibfnamefont {B.~V.}\ \bibnamefont
  {Lehmann}}, \bibinfo {author} {\bibfnamefont {I.}~\bibnamefont {Charaev}},
  \bibinfo {author} {\bibfnamefont {J.}~\bibnamefont {Chiles}}, \bibinfo
  {author} {\bibfnamefont {M.}~\bibnamefont {Colangelo}}, \bibinfo {author}
  {\bibfnamefont {S.~W.}\ \bibnamefont {Nam}},\ and\ \bibinfo {author}
  {\bibfnamefont {K.~K.}\ \bibnamefont {Berggren}},\ }\bibfield  {title}
  {\bibinfo {title} {{New Constraints on Dark Matter from Superconducting
  Nanowires}},\ }\href@noop {} {\  (\bibinfo {year} {2021}{\natexlab{a}})},\
  \Eprint {https://arxiv.org/abs/2110.01586} {arXiv:2110.01586 [hep-ph]}
  \BibitemShut {NoStop}%
\bibitem [{\citenamefont {Chiles}\ \emph {et~al.}(2021)\citenamefont {Chiles}
  \emph {et~al.}}]{Chiles:2021gxk}%
  \BibitemOpen
  \bibfield  {author} {\bibinfo {author} {\bibfnamefont {J.}~\bibnamefont
  {Chiles}} \emph {et~al.},\ }\bibfield  {title} {\bibinfo {title} {{First
  Constraints on Dark Photon Dark Matter with Superconducting Nanowire
  Detectors in an Optical Haloscope}},\ }\href@noop {} {\  (\bibinfo {year}
  {2021})},\ \Eprint {https://arxiv.org/abs/2110.01582} {arXiv:2110.01582
  [hep-ex]} \BibitemShut {NoStop}%
\bibitem [{\citenamefont {{D.\ N.\ McKinsey {\it et al.} [TESSERACT
  Collaboration]}}()}]{SPICE}%
  \BibitemOpen
  \bibfield  {author} {\bibinfo {author} {\bibnamefont {{D.\ N.\ McKinsey {\it
  et al.} [TESSERACT Collaboration]}}} (\bibinfo {collaboration} {TESSERACT}),\
  }\href
  {www.snowmass21.org/docs/files/summaries/CF/SNOWMASS21-CF1_CF2-IF1_IF8-120.pdf}
  {\bibinfo {title} {{The TESSERACT Dark Matter Project, SNOWMASS LOI}}},\
  \bibinfo {howpublished} {Availabe
  \href{www.snowmass21.org/docs/files/summaries/CF/SNOWMASS21-CF1_CF2-IF1_IF8-120.pdf}{[Online]}}\BibitemShut
  {NoStop}%
\bibitem [{\citenamefont {Essig}\ \emph
  {et~al.}(2012{\natexlab{b}})\citenamefont {Essig}, \citenamefont {Mardon},\
  and\ \citenamefont {Volansky}}]{Essig:2011nj}%
  \BibitemOpen
  \bibfield  {author} {\bibinfo {author} {\bibfnamefont {R.}~\bibnamefont
  {Essig}}, \bibinfo {author} {\bibfnamefont {J.}~\bibnamefont {Mardon}},\ and\
  \bibinfo {author} {\bibfnamefont {T.}~\bibnamefont {Volansky}},\ }\bibfield
  {title} {\bibinfo {title} {{Direct Detection of Sub-GeV Dark Matter}},\
  }\href {https://doi.org/10.1103/PhysRevD.85.076007} {\bibfield  {journal}
  {\bibinfo  {journal} {Phys. Rev. D}\ }\textbf {\bibinfo {volume} {85}},\
  \bibinfo {pages} {076007} (\bibinfo {year} {2012}{\natexlab{b}})},\ \Eprint
  {https://arxiv.org/abs/1108.5383} {arXiv:1108.5383 [hep-ph]} \BibitemShut
  {NoStop}%
\bibitem [{\citenamefont {Graham}\ \emph {et~al.}(2012)\citenamefont {Graham},
  \citenamefont {Kaplan}, \citenamefont {Rajendran},\ and\ \citenamefont
  {Walters}}]{Graham:2012su}%
  \BibitemOpen
  \bibfield  {author} {\bibinfo {author} {\bibfnamefont {P.~W.}\ \bibnamefont
  {Graham}}, \bibinfo {author} {\bibfnamefont {D.~E.}\ \bibnamefont {Kaplan}},
  \bibinfo {author} {\bibfnamefont {S.}~\bibnamefont {Rajendran}},\ and\
  \bibinfo {author} {\bibfnamefont {M.~T.}\ \bibnamefont {Walters}},\
  }\bibfield  {title} {\bibinfo {title} {{Semiconductor Probes of Light Dark
  Matter}},\ }\href {https://doi.org/10.1016/j.dark.2012.09.001} {\bibfield
  {journal} {\bibinfo  {journal} {Phys. Dark Univ.}\ }\textbf {\bibinfo
  {volume} {1}},\ \bibinfo {pages} {32} (\bibinfo {year} {2012})},\ \Eprint
  {https://arxiv.org/abs/1203.2531} {arXiv:1203.2531 [hep-ph]} \BibitemShut
  {NoStop}%
\bibitem [{\citenamefont {Essig}\ \emph {et~al.}(2016)\citenamefont {Essig},
  \citenamefont {Fernandez-Serra}, \citenamefont {Mardon}, \citenamefont
  {Soto}, \citenamefont {Volansky},\ and\ \citenamefont {Yu}}]{Essig:2015cda}%
  \BibitemOpen
  \bibfield  {author} {\bibinfo {author} {\bibfnamefont {R.}~\bibnamefont
  {Essig}}, \bibinfo {author} {\bibfnamefont {M.}~\bibnamefont
  {Fernandez-Serra}}, \bibinfo {author} {\bibfnamefont {J.}~\bibnamefont
  {Mardon}}, \bibinfo {author} {\bibfnamefont {A.}~\bibnamefont {Soto}},
  \bibinfo {author} {\bibfnamefont {T.}~\bibnamefont {Volansky}},\ and\
  \bibinfo {author} {\bibfnamefont {T.-T.}\ \bibnamefont {Yu}},\ }\bibfield
  {title} {\bibinfo {title} {{Direct Detection of Sub-Gev Dark Matter with
  Semiconductor Targets}},\ }\href {https://doi.org/10.1007/JHEP05(2016)046}
  {\bibfield  {journal} {\bibinfo  {journal} {JHEP}\ }\textbf {\bibinfo
  {volume} {05}},\ \bibinfo {pages} {046}},\ \Eprint
  {https://arxiv.org/abs/1509.01598} {arXiv:1509.01598 [hep-ph]} \BibitemShut
  {NoStop}%
\bibitem [{\citenamefont {Lee}\ \emph {et~al.}(2015)\citenamefont {Lee},
  \citenamefont {Lisanti}, \citenamefont {Mishra-Sharma},\ and\ \citenamefont
  {Safdi}}]{Lee:2015qva}%
  \BibitemOpen
  \bibfield  {author} {\bibinfo {author} {\bibfnamefont {S.~K.}\ \bibnamefont
  {Lee}}, \bibinfo {author} {\bibfnamefont {M.}~\bibnamefont {Lisanti}},
  \bibinfo {author} {\bibfnamefont {S.}~\bibnamefont {Mishra-Sharma}},\ and\
  \bibinfo {author} {\bibfnamefont {B.~R.}\ \bibnamefont {Safdi}},\ }\bibfield
  {title} {\bibinfo {title} {{Modulation Effects in Dark Matter-Electron
  Scattering Experiments}},\ }\href
  {https://doi.org/10.1103/PhysRevD.92.083517} {\bibfield  {journal} {\bibinfo
  {journal} {Phys. Rev.}\ }\textbf {\bibinfo {volume} {D92}},\ \bibinfo {pages}
  {083517} (\bibinfo {year} {2015})},\ \Eprint
  {https://arxiv.org/abs/1508.07361} {arXiv:1508.07361 [hep-ph]} \BibitemShut
  {NoStop}%
\bibitem [{\citenamefont {Hochberg}\ \emph {et~al.}(2017)\citenamefont
  {Hochberg}, \citenamefont {Kahn}, \citenamefont {Lisanti}, \citenamefont
  {Tully},\ and\ \citenamefont {Zurek}}]{Hochberg:2016ntt}%
  \BibitemOpen
  \bibfield  {author} {\bibinfo {author} {\bibfnamefont {Y.}~\bibnamefont
  {Hochberg}}, \bibinfo {author} {\bibfnamefont {Y.}~\bibnamefont {Kahn}},
  \bibinfo {author} {\bibfnamefont {M.}~\bibnamefont {Lisanti}}, \bibinfo
  {author} {\bibfnamefont {C.~G.}\ \bibnamefont {Tully}},\ and\ \bibinfo
  {author} {\bibfnamefont {K.~M.}\ \bibnamefont {Zurek}},\ }\bibfield  {title}
  {\bibinfo {title} {{Directional Detection of Dark Matter with Two-Dimensional
  Targets}},\ }\href {https://doi.org/10.1016/j.physletb.2017.06.051}
  {\bibfield  {journal} {\bibinfo  {journal} {Phys. Lett.}\ }\textbf {\bibinfo
  {volume} {B772}},\ \bibinfo {pages} {239} (\bibinfo {year} {2017})},\ \Eprint
  {https://arxiv.org/abs/1606.08849} {arXiv:1606.08849 [hep-ph]} \BibitemShut
  {NoStop}%
\bibitem [{\citenamefont {Derenzo}\ \emph {et~al.}(2017)\citenamefont
  {Derenzo}, \citenamefont {Essig}, \citenamefont {Massari}, \citenamefont
  {Soto},\ and\ \citenamefont {Yu}}]{Derenzo:2016fse}%
  \BibitemOpen
  \bibfield  {author} {\bibinfo {author} {\bibfnamefont {S.}~\bibnamefont
  {Derenzo}}, \bibinfo {author} {\bibfnamefont {R.}~\bibnamefont {Essig}},
  \bibinfo {author} {\bibfnamefont {A.}~\bibnamefont {Massari}}, \bibinfo
  {author} {\bibfnamefont {A.}~\bibnamefont {Soto}},\ and\ \bibinfo {author}
  {\bibfnamefont {T.-T.}\ \bibnamefont {Yu}},\ }\bibfield  {title} {\bibinfo
  {title} {{Direct Detection of sub-GeV Dark Matter with Scintillating
  Targets}},\ }\href {https://doi.org/10.1103/PhysRevD.96.016026} {\bibfield
  {journal} {\bibinfo  {journal} {Phys. Rev.}\ }\textbf {\bibinfo {volume}
  {D96}},\ \bibinfo {pages} {016026} (\bibinfo {year} {2017})},\ \Eprint
  {https://arxiv.org/abs/1607.01009} {arXiv:1607.01009 [hep-ph]} \BibitemShut
  {NoStop}%
\bibitem [{\citenamefont {Knapen}\ \emph {et~al.}(2018)\citenamefont {Knapen},
  \citenamefont {Lin}, \citenamefont {Pyle},\ and\ \citenamefont
  {Zurek}}]{Knapen:2017ekk}%
  \BibitemOpen
  \bibfield  {author} {\bibinfo {author} {\bibfnamefont {S.}~\bibnamefont
  {Knapen}}, \bibinfo {author} {\bibfnamefont {T.}~\bibnamefont {Lin}},
  \bibinfo {author} {\bibfnamefont {M.}~\bibnamefont {Pyle}},\ and\ \bibinfo
  {author} {\bibfnamefont {K.~M.}\ \bibnamefont {Zurek}},\ }\bibfield  {title}
  {\bibinfo {title} {{Detection of Light Dark Matter With Optical Phonons in
  Polar Materials}},\ }\href {https://doi.org/10.1016/j.physletb.2018.08.064}
  {\bibfield  {journal} {\bibinfo  {journal} {Phys. Lett. B}\ }\textbf
  {\bibinfo {volume} {785}},\ \bibinfo {pages} {386} (\bibinfo {year}
  {2018})},\ \Eprint {https://arxiv.org/abs/1712.06598} {arXiv:1712.06598
  [hep-ph]} \BibitemShut {NoStop}%
\bibitem [{\citenamefont {Griffin}\ \emph {et~al.}(2018)\citenamefont
  {Griffin}, \citenamefont {Knapen}, \citenamefont {Lin},\ and\ \citenamefont
  {Zurek}}]{Griffin:2018bjn}%
  \BibitemOpen
  \bibfield  {author} {\bibinfo {author} {\bibfnamefont {S.}~\bibnamefont
  {Griffin}}, \bibinfo {author} {\bibfnamefont {S.}~\bibnamefont {Knapen}},
  \bibinfo {author} {\bibfnamefont {T.}~\bibnamefont {Lin}},\ and\ \bibinfo
  {author} {\bibfnamefont {K.~M.}\ \bibnamefont {Zurek}},\ }\bibfield  {title}
  {\bibinfo {title} {{Directional Detection of Light Dark Matter with Polar
  Materials}},\ }\href {https://doi.org/10.1103/PhysRevD.98.115034} {\bibfield
  {journal} {\bibinfo  {journal} {Phys. Rev. D}\ }\textbf {\bibinfo {volume}
  {98}},\ \bibinfo {pages} {115034} (\bibinfo {year} {2018})},\ \Eprint
  {https://arxiv.org/abs/1807.10291} {arXiv:1807.10291 [hep-ph]} \BibitemShut
  {NoStop}%
\bibitem [{\citenamefont {Kurinsky}\ \emph {et~al.}(2019)\citenamefont
  {Kurinsky}, \citenamefont {Yu}, \citenamefont {Hochberg},\ and\ \citenamefont
  {Cabrera}}]{Kurinsky:2019pgb}%
  \BibitemOpen
  \bibfield  {author} {\bibinfo {author} {\bibfnamefont {N.~A.}\ \bibnamefont
  {Kurinsky}}, \bibinfo {author} {\bibfnamefont {T.~C.}\ \bibnamefont {Yu}},
  \bibinfo {author} {\bibfnamefont {Y.}~\bibnamefont {Hochberg}},\ and\
  \bibinfo {author} {\bibfnamefont {B.}~\bibnamefont {Cabrera}},\ }\bibfield
  {title} {\bibinfo {title} {{Diamond Detectors for Direct Detection of Sub-GeV
  Dark Matter}},\ }\href {https://doi.org/10.1103/PhysRevD.99.123005}
  {\bibfield  {journal} {\bibinfo  {journal} {Phys. Rev. D}\ }\textbf {\bibinfo
  {volume} {99}},\ \bibinfo {pages} {123005} (\bibinfo {year} {2019})},\
  \Eprint {https://arxiv.org/abs/1901.07569} {arXiv:1901.07569 [hep-ex]}
  \BibitemShut {NoStop}%
\bibitem [{\citenamefont {Griffin}\ \emph {et~al.}(2020)\citenamefont
  {Griffin}, \citenamefont {Inzani}, \citenamefont {Trickle}, \citenamefont
  {Zhang},\ and\ \citenamefont {Zurek}}]{Griffin:2019mvc}%
  \BibitemOpen
  \bibfield  {author} {\bibinfo {author} {\bibfnamefont {S.~M.}\ \bibnamefont
  {Griffin}}, \bibinfo {author} {\bibfnamefont {K.}~\bibnamefont {Inzani}},
  \bibinfo {author} {\bibfnamefont {T.}~\bibnamefont {Trickle}}, \bibinfo
  {author} {\bibfnamefont {Z.}~\bibnamefont {Zhang}},\ and\ \bibinfo {author}
  {\bibfnamefont {K.~M.}\ \bibnamefont {Zurek}},\ }\bibfield  {title} {\bibinfo
  {title} {{Multichannel direct detection of light dark matter: Target
  comparison}},\ }\href {https://doi.org/10.1103/PhysRevD.101.055004}
  {\bibfield  {journal} {\bibinfo  {journal} {Phys. Rev. D}\ }\textbf {\bibinfo
  {volume} {101}},\ \bibinfo {pages} {055004} (\bibinfo {year} {2020})},\
  \Eprint {https://arxiv.org/abs/1910.10716} {arXiv:1910.10716 [hep-ph]}
  \BibitemShut {NoStop}%
\bibitem [{\citenamefont {Caputo}\ \emph {et~al.}(2019)\citenamefont {Caputo},
  \citenamefont {Esposito},\ and\ \citenamefont {Polosa}}]{Caputo:2019cyg}%
  \BibitemOpen
  \bibfield  {author} {\bibinfo {author} {\bibfnamefont {A.}~\bibnamefont
  {Caputo}}, \bibinfo {author} {\bibfnamefont {A.}~\bibnamefont {Esposito}},\
  and\ \bibinfo {author} {\bibfnamefont {A.~D.}\ \bibnamefont {Polosa}},\
  }\bibfield  {title} {\bibinfo {title} {{Sub-MeV Dark Matter and the Goldstone
  Modes of Superfluid Helium}},\ }\href
  {https://doi.org/10.1103/PhysRevD.100.116007} {\bibfield  {journal} {\bibinfo
   {journal} {Phys. Rev. D}\ }\textbf {\bibinfo {volume} {100}},\ \bibinfo
  {pages} {116007} (\bibinfo {year} {2019})},\ \Eprint
  {https://arxiv.org/abs/1907.10635} {arXiv:1907.10635 [hep-ph]} \BibitemShut
  {NoStop}%
\bibitem [{\citenamefont {Trickle}\ \emph
  {et~al.}(2020{\natexlab{a}})\citenamefont {Trickle}, \citenamefont {Zhang},
  \citenamefont {Zurek}, \citenamefont {Inzani},\ and\ \citenamefont
  {Griffin}}]{Trickle:2019nya}%
  \BibitemOpen
  \bibfield  {author} {\bibinfo {author} {\bibfnamefont {T.}~\bibnamefont
  {Trickle}}, \bibinfo {author} {\bibfnamefont {Z.}~\bibnamefont {Zhang}},
  \bibinfo {author} {\bibfnamefont {K.~M.}\ \bibnamefont {Zurek}}, \bibinfo
  {author} {\bibfnamefont {K.}~\bibnamefont {Inzani}},\ and\ \bibinfo {author}
  {\bibfnamefont {S.}~\bibnamefont {Griffin}},\ }\bibfield  {title} {\bibinfo
  {title} {{Multi-Channel Direct Detection of Light Dark Matter: Theoretical
  Framework}},\ }\href {https://doi.org/10.1007/JHEP03(2020)036} {\bibfield
  {journal} {\bibinfo  {journal} {JHEP}\ }\textbf {\bibinfo {volume} {03}},\
  \bibinfo {pages} {036}},\ \Eprint {https://arxiv.org/abs/1910.08092}
  {arXiv:1910.08092 [hep-ph]} \BibitemShut {NoStop}%
\bibitem [{\citenamefont {Trickle}\ \emph {et~al.}(2022)\citenamefont
  {Trickle}, \citenamefont {Zhang},\ and\ \citenamefont
  {Zurek}}]{Trickle:2020oki}%
  \BibitemOpen
  \bibfield  {author} {\bibinfo {author} {\bibfnamefont {T.}~\bibnamefont
  {Trickle}}, \bibinfo {author} {\bibfnamefont {Z.}~\bibnamefont {Zhang}},\
  and\ \bibinfo {author} {\bibfnamefont {K.~M.}\ \bibnamefont {Zurek}},\
  }\bibfield  {title} {\bibinfo {title} {{Effective field theory of dark matter
  direct detection with collective excitations}},\ }\href
  {https://doi.org/10.1103/PhysRevD.105.015001} {\bibfield  {journal} {\bibinfo
   {journal} {Phys. Rev. D}\ }\textbf {\bibinfo {volume} {105}},\ \bibinfo
  {pages} {015001} (\bibinfo {year} {2022})},\ \Eprint
  {https://arxiv.org/abs/2009.13534} {arXiv:2009.13534 [hep-ph]} \BibitemShut
  {NoStop}%
\bibitem [{\citenamefont {Griffin}\ \emph
  {et~al.}(2021{\natexlab{a}})\citenamefont {Griffin}, \citenamefont
  {Hochberg}, \citenamefont {Inzani}, \citenamefont {Kurinsky}, \citenamefont
  {Lin},\ and\ \citenamefont {Chin}}]{Griffin:2020lgd}%
  \BibitemOpen
  \bibfield  {author} {\bibinfo {author} {\bibfnamefont {S.~M.}\ \bibnamefont
  {Griffin}}, \bibinfo {author} {\bibfnamefont {Y.}~\bibnamefont {Hochberg}},
  \bibinfo {author} {\bibfnamefont {K.}~\bibnamefont {Inzani}}, \bibinfo
  {author} {\bibfnamefont {N.}~\bibnamefont {Kurinsky}}, \bibinfo {author}
  {\bibfnamefont {T.}~\bibnamefont {Lin}},\ and\ \bibinfo {author}
  {\bibfnamefont {T.}~\bibnamefont {Chin}},\ }\bibfield  {title} {\bibinfo
  {title} {{Silicon carbide detectors for sub-GeV dark matter}},\ }\href
  {https://doi.org/10.1103/PhysRevD.103.075002} {\bibfield  {journal} {\bibinfo
   {journal} {Phys. Rev. D}\ }\textbf {\bibinfo {volume} {103}},\ \bibinfo
  {pages} {075002} (\bibinfo {year} {2021}{\natexlab{a}})},\ \Eprint
  {https://arxiv.org/abs/2008.08560} {arXiv:2008.08560 [hep-ph]} \BibitemShut
  {NoStop}%
\bibitem [{\citenamefont {Griffin}\ \emph
  {et~al.}(2021{\natexlab{b}})\citenamefont {Griffin}, \citenamefont {Inzani},
  \citenamefont {Trickle}, \citenamefont {Zhang},\ and\ \citenamefont
  {Zurek}}]{Griffin:2021znd}%
  \BibitemOpen
  \bibfield  {author} {\bibinfo {author} {\bibfnamefont {S.~M.}\ \bibnamefont
  {Griffin}}, \bibinfo {author} {\bibfnamefont {K.}~\bibnamefont {Inzani}},
  \bibinfo {author} {\bibfnamefont {T.}~\bibnamefont {Trickle}}, \bibinfo
  {author} {\bibfnamefont {Z.}~\bibnamefont {Zhang}},\ and\ \bibinfo {author}
  {\bibfnamefont {K.~M.}\ \bibnamefont {Zurek}},\ }\bibfield  {title} {\bibinfo
  {title} {{Extended calculation of dark matter-electron scattering in crystal
  targets}},\ }\href {https://doi.org/10.1103/PhysRevD.104.095015} {\bibfield
  {journal} {\bibinfo  {journal} {Phys. Rev. D}\ }\textbf {\bibinfo {volume}
  {104}},\ \bibinfo {pages} {095015} (\bibinfo {year} {2021}{\natexlab{b}})},\
  \Eprint {https://arxiv.org/abs/2105.05253} {arXiv:2105.05253 [hep-ph]}
  \BibitemShut {NoStop}%
\bibitem [{\citenamefont {Coskuner}\ \emph {et~al.}(2022)\citenamefont
  {Coskuner}, \citenamefont {Trickle}, \citenamefont {Zhang},\ and\
  \citenamefont {Zurek}}]{Coskuner:2021qxo}%
  \BibitemOpen
  \bibfield  {author} {\bibinfo {author} {\bibfnamefont {A.}~\bibnamefont
  {Coskuner}}, \bibinfo {author} {\bibfnamefont {T.}~\bibnamefont {Trickle}},
  \bibinfo {author} {\bibfnamefont {Z.}~\bibnamefont {Zhang}},\ and\ \bibinfo
  {author} {\bibfnamefont {K.~M.}\ \bibnamefont {Zurek}},\ }\bibfield  {title}
  {\bibinfo {title} {{Directional detectability of dark matter with single
  phonon excitations: Target comparison}},\ }\href
  {https://doi.org/10.1103/PhysRevD.105.015010} {\bibfield  {journal} {\bibinfo
   {journal} {Phys. Rev. D}\ }\textbf {\bibinfo {volume} {105}},\ \bibinfo
  {pages} {015010} (\bibinfo {year} {2022})},\ \Eprint
  {https://arxiv.org/abs/2102.09567} {arXiv:2102.09567 [hep-ph]} \BibitemShut
  {NoStop}%
\bibitem [{\citenamefont {Robinson}(2017)}]{Robinson_2017}%
  \BibitemOpen
  \bibfield  {author} {\bibinfo {author} {\bibfnamefont {A.~E.}\ \bibnamefont
  {Robinson}},\ }\bibfield  {title} {\bibinfo {title} {Erratum: Coherent photon
  scattering background in sub- gev/c2 direct dark matter searches [phys. rev.
  d 95 , 021301(r) (2017)]},\ }\bibfield  {journal} {\bibinfo  {journal}
  {Physical Review D}\ }\textbf {\bibinfo {volume} {95}},\ \href
  {https://doi.org/10.1103/physrevd.95.069907} {10.1103/physrevd.95.069907}
  (\bibinfo {year} {2017})\BibitemShut {NoStop}%
\bibitem [{\citenamefont {Du}\ \emph {et~al.}(2020)\citenamefont {Du},
  \citenamefont {Egana-Ugrinovic}, \citenamefont {Essig},\ and\ \citenamefont
  {Sholapurkar}}]{Du:2020ldo}%
  \BibitemOpen
  \bibfield  {author} {\bibinfo {author} {\bibfnamefont {P.}~\bibnamefont
  {Du}}, \bibinfo {author} {\bibfnamefont {D.}~\bibnamefont {Egana-Ugrinovic}},
  \bibinfo {author} {\bibfnamefont {R.}~\bibnamefont {Essig}},\ and\ \bibinfo
  {author} {\bibfnamefont {M.}~\bibnamefont {Sholapurkar}},\ }\bibfield
  {title} {\bibinfo {title} {{Sources of Low-Energy Events in Low-Threshold
  Dark Matter Detectors}},\ }\href@noop {} {\  (\bibinfo {year} {2020})},\
  \Eprint {https://arxiv.org/abs/2011.13939} {arXiv:2011.13939 [hep-ph]}
  \BibitemShut {NoStop}%
\bibitem [{\citenamefont {Robinson}(2020)}]{Robinson:2020zec}%
  \BibitemOpen
  \bibfield  {author} {\bibinfo {author} {\bibfnamefont {A.~E.}\ \bibnamefont
  {Robinson}},\ }\bibfield  {title} {\bibinfo {title} {{Electrovolt-scale
  backgrounds from surfaces}},\ }\href@noop {} {\  (\bibinfo {year} {2020})},\
  \Eprint {https://arxiv.org/abs/2010.11043} {arXiv:2010.11043 [astro-ph.IM]}
  \BibitemShut {NoStop}%
\bibitem [{\citenamefont {Kane}\ \emph {et~al.}(1986)\citenamefont {Kane},
  \citenamefont {Kissel}, \citenamefont {Pratt},\ and\ \citenamefont
  {Roy}}]{KANE198675}%
  \BibitemOpen
  \bibfield  {author} {\bibinfo {author} {\bibfnamefont {P.}~\bibnamefont
  {Kane}}, \bibinfo {author} {\bibfnamefont {L.}~\bibnamefont {Kissel}},
  \bibinfo {author} {\bibfnamefont {R.}~\bibnamefont {Pratt}},\ and\ \bibinfo
  {author} {\bibfnamefont {S.}~\bibnamefont {Roy}},\ }\bibfield  {title}
  {\bibinfo {title} {Elastic scattering of $\gamma$-rays and x-rays by atoms},\
  }\href {https://doi.org/https://doi.org/10.1016/0370-1573(86)90018-9}
  {\bibfield  {journal} {\bibinfo  {journal} {Physics Reports}\ }\textbf
  {\bibinfo {volume} {140}},\ \bibinfo {pages} {75} (\bibinfo {year}
  {1986})}\BibitemShut {NoStop}%
\bibitem [{\citenamefont {Chatterjee}\ and\ \citenamefont
  {Roy}(1998)}]{doi:10.1063/1.556027}%
  \BibitemOpen
  \bibfield  {author} {\bibinfo {author} {\bibfnamefont {B.~K.}\ \bibnamefont
  {Chatterjee}}\ and\ \bibinfo {author} {\bibfnamefont {S.~C.}\ \bibnamefont
  {Roy}},\ }\bibfield  {title} {\bibinfo {title} {Tables of elastic scattering
  cross sections of photons in the energy range 50–1500 kev for all elements
  in the range $13 \le z \le 104$},\ }\href {https://doi.org/10.1063/1.556027}
  {\bibfield  {journal} {\bibinfo  {journal} {Journal of Physical and Chemical
  Reference Data}\ }\textbf {\bibinfo {volume} {27}},\ \bibinfo {pages} {1011}
  (\bibinfo {year} {1998})},\ \Eprint
  {https://arxiv.org/abs/https://doi.org/10.1063/1.556027}
  {https://doi.org/10.1063/1.556027} \BibitemShut {NoStop}%
\bibitem [{\citenamefont {Roy}\ \emph {et~al.}(1999)\citenamefont {Roy},
  \citenamefont {Kissel},\ and\ \citenamefont {Pratt}}]{ROY19993}%
  \BibitemOpen
  \bibfield  {author} {\bibinfo {author} {\bibfnamefont {S.}~\bibnamefont
  {Roy}}, \bibinfo {author} {\bibfnamefont {L.}~\bibnamefont {Kissel}},\ and\
  \bibinfo {author} {\bibfnamefont {R.}~\bibnamefont {Pratt}},\ }\bibfield
  {title} {\bibinfo {title} {Elastic scattering of photons},\ }\href
  {https://doi.org/https://doi.org/10.1016/S0969-806X(99)00286-8} {\bibfield
  {journal} {\bibinfo  {journal} {Radiation Physics and Chemistry}\ }\textbf
  {\bibinfo {volume} {56}},\ \bibinfo {pages} {3} (\bibinfo {year}
  {1999})}\BibitemShut {NoStop}%
\bibitem [{\citenamefont {Kissel}\ \emph {et~al.}(1995)\citenamefont {Kissel},
  \citenamefont {Zhou}, \citenamefont {Roy}, \citenamefont {Sen~Gupta},\ and\
  \citenamefont {Pratt}}]{Kissel:st0621}%
  \BibitemOpen
  \bibfield  {author} {\bibinfo {author} {\bibfnamefont {L.}~\bibnamefont
  {Kissel}}, \bibinfo {author} {\bibfnamefont {B.}~\bibnamefont {Zhou}},
  \bibinfo {author} {\bibfnamefont {S.~C.}\ \bibnamefont {Roy}}, \bibinfo
  {author} {\bibfnamefont {S.~K.}\ \bibnamefont {Sen~Gupta}},\ and\ \bibinfo
  {author} {\bibfnamefont {R.~H.}\ \bibnamefont {Pratt}},\ }\bibfield  {title}
  {\bibinfo {title} {{The validity of form-factor, modified-form-factor and
  anomalous-scattering-factor approximations in elastic scattering
  calculations}},\ }\href {https://doi.org/10.1107/S010876739400886X}
  {\bibfield  {journal} {\bibinfo  {journal} {Acta Crystallographica Section
  A}\ }\textbf {\bibinfo {volume} {51}},\ \bibinfo {pages} {271} (\bibinfo
  {year} {1995})}\BibitemShut {NoStop}%
\bibitem [{\citenamefont {Gu}(2008)}]{doi:10.1139/p07-197}%
  \BibitemOpen
  \bibfield  {author} {\bibinfo {author} {\bibfnamefont {M.~F.}\ \bibnamefont
  {Gu}},\ }\bibfield  {title} {\bibinfo {title} {The flexible atomic code},\
  }\href {https://doi.org/10.1139/p07-197} {\bibfield  {journal} {\bibinfo
  {journal} {Canadian Journal of Physics}\ }\textbf {\bibinfo {volume} {86}},\
  \bibinfo {pages} {675} (\bibinfo {year} {2008})},\ \Eprint
  {https://arxiv.org/abs/https://doi.org/10.1139/p07-197}
  {https://doi.org/10.1139/p07-197} \BibitemShut {NoStop}%
\bibitem [{\citenamefont {Wilson}\ and\ \citenamefont
  {Geist}(1993)}]{https://doi.org/10.1002/crat.2170280117}%
  \BibitemOpen
  \bibfield  {author} {\bibinfo {author} {\bibfnamefont {A.~J.~C.}\
  \bibnamefont {Wilson}}\ and\ \bibinfo {author} {\bibfnamefont
  {V.}~\bibnamefont {Geist}},\ }\bibfield  {title} {\bibinfo {title}
  {International tables for crystallography. volume c: Mathematical, physical
  and chemical tables. kluwer academic publishers, dordrecht/boston/london 1992
  (published for the international union of crystallography), 883 seiten, isbn
  0-792-3-16-38x},\ }\href
  {https://doi.org/https://doi.org/10.1002/crat.2170280117} {\bibfield
  {journal} {\bibinfo  {journal} {Crystal Research and Technology}\ }\textbf
  {\bibinfo {volume} {28}},\ \bibinfo {pages} {110} (\bibinfo {year} {1993})},\
  \Eprint
  {https://arxiv.org/abs/https://onlinelibrary.wiley.com/doi/pdf/10.1002/crat.2170280117}
  {https://onlinelibrary.wiley.com/doi/pdf/10.1002/crat.2170280117}
  \BibitemShut {NoStop}%
\bibitem [{\citenamefont {Falkenberg}\ \emph {et~al.}(1992)\citenamefont
  {Falkenberg}, \citenamefont {Hünger}, \citenamefont {Rullhusen},
  \citenamefont {Schumacher}, \citenamefont {Milstein},\ and\ \citenamefont
  {Mork}}]{FALKENBERG19921}%
  \BibitemOpen
  \bibfield  {author} {\bibinfo {author} {\bibfnamefont {H.}~\bibnamefont
  {Falkenberg}}, \bibinfo {author} {\bibfnamefont {A.}~\bibnamefont {Hünger}},
  \bibinfo {author} {\bibfnamefont {P.}~\bibnamefont {Rullhusen}}, \bibinfo
  {author} {\bibfnamefont {M.}~\bibnamefont {Schumacher}}, \bibinfo {author}
  {\bibfnamefont {A.}~\bibnamefont {Milstein}},\ and\ \bibinfo {author}
  {\bibfnamefont {K.}~\bibnamefont {Mork}},\ }\bibfield  {title} {\bibinfo
  {title} {Amplitudes for delbrück scattering},\ }\href
  {https://doi.org/https://doi.org/10.1016/0092-640X(92)90023-B} {\bibfield
  {journal} {\bibinfo  {journal} {Atomic Data and Nuclear Data Tables}\
  }\textbf {\bibinfo {volume} {50}},\ \bibinfo {pages} {1} (\bibinfo {year}
  {1992})}\BibitemShut {NoStop}%
\bibitem [{\citenamefont {Rullhusen}\ \emph {et~al.}(1981)\citenamefont
  {Rullhusen}, \citenamefont {M\"uckenheim}, \citenamefont {Smend},
  \citenamefont {Schumacher}, \citenamefont {Berg}, \citenamefont {Mork},\ and\
  \citenamefont {Kissel}}]{PhysRevC.23.1375}%
  \BibitemOpen
  \bibfield  {author} {\bibinfo {author} {\bibfnamefont {P.}~\bibnamefont
  {Rullhusen}}, \bibinfo {author} {\bibfnamefont {W.}~\bibnamefont
  {M\"uckenheim}}, \bibinfo {author} {\bibfnamefont {F.}~\bibnamefont {Smend}},
  \bibinfo {author} {\bibfnamefont {M.}~\bibnamefont {Schumacher}}, \bibinfo
  {author} {\bibfnamefont {G.~P.~A.}\ \bibnamefont {Berg}}, \bibinfo {author}
  {\bibfnamefont {K.}~\bibnamefont {Mork}},\ and\ \bibinfo {author}
  {\bibfnamefont {L.}~\bibnamefont {Kissel}},\ }\bibfield  {title} {\bibinfo
  {title} {Test of vacuum polarization by precise investigation of delbr\"uck
  scattering},\ }\href {https://doi.org/10.1103/PhysRevC.23.1375} {\bibfield
  {journal} {\bibinfo  {journal} {Phys. Rev. C}\ }\textbf {\bibinfo {volume}
  {23}},\ \bibinfo {pages} {1375} (\bibinfo {year} {1981})}\BibitemShut
  {NoStop}%
\bibitem [{\citenamefont {Rullhusen}\ \emph {et~al.}(1979)\citenamefont
  {Rullhusen}, \citenamefont {Smend},\ and\ \citenamefont
  {Schumacher}}]{Rullhusen:1979zza}%
  \BibitemOpen
  \bibfield  {author} {\bibinfo {author} {\bibfnamefont {P.}~\bibnamefont
  {Rullhusen}}, \bibinfo {author} {\bibfnamefont {F.}~\bibnamefont {Smend}},\
  and\ \bibinfo {author} {\bibfnamefont {M.}~\bibnamefont {Schumacher}},\
  }\bibfield  {title} {\bibinfo {title} {{Delbruck Scattering of 2.754 MeV
  photons by Nd, Ce, I. Sn, Mo and Zn}},\ }\href
  {https://doi.org/10.1016/0370-2693(79)90274-0} {\bibfield  {journal}
  {\bibinfo  {journal} {Phys. Lett. B}\ }\textbf {\bibinfo {volume} {84}},\
  \bibinfo {pages} {166} (\bibinfo {year} {1979})}\BibitemShut {NoStop}%
\bibitem [{\citenamefont {Rullhusen}\ and\ \citenamefont
  {Schumacher}(1979)}]{Rullhusen:1979zz}%
  \BibitemOpen
  \bibfield  {author} {\bibinfo {author} {\bibfnamefont {P.}~\bibnamefont
  {Rullhusen}}\ and\ \bibinfo {author} {\bibfnamefont {M.}~\bibnamefont
  {Schumacher}},\ }\bibfield  {title} {\bibinfo {title} {{Coulomb Correction to
  Delbruck Scattering Investigated at Z=94}},\ }\href
  {https://doi.org/10.1007/BF01435270} {\bibfield  {journal} {\bibinfo
  {journal} {Z. Phys. A}\ }\textbf {\bibinfo {volume} {293}},\ \bibinfo {pages}
  {287} (\bibinfo {year} {1979})}\BibitemShut {NoStop}%
\bibitem [{\citenamefont {Berman}\ and\ \citenamefont
  {Fultz}(1975)}]{Berman:1975tt}%
  \BibitemOpen
  \bibfield  {author} {\bibinfo {author} {\bibfnamefont {B.~L.}\ \bibnamefont
  {Berman}}\ and\ \bibinfo {author} {\bibfnamefont {S.~C.}\ \bibnamefont
  {Fultz}},\ }\bibfield  {title} {\bibinfo {title} {{Measurements of the giant
  dipole resonance with monoenergetic photons}},\ }\href
  {https://doi.org/10.1103/RevModPhys.47.713} {\bibfield  {journal} {\bibinfo
  {journal} {Rev. Mod. Phys.}\ }\textbf {\bibinfo {volume} {47}},\ \bibinfo
  {pages} {713} (\bibinfo {year} {1975})}\BibitemShut {NoStop}%
\bibitem [{\citenamefont {Kramers}\ and\ \citenamefont
  {Heisenberg}(1925)}]{Kramers:1925ucm}%
  \BibitemOpen
  \bibfield  {author} {\bibinfo {author} {\bibfnamefont {H.~A.}\ \bibnamefont
  {Kramers}}\ and\ \bibinfo {author} {\bibfnamefont {W.}~\bibnamefont
  {Heisenberg}},\ }\bibfield  {title} {\bibinfo {title} {{\"Uber die Streuung
  von Strahlung durch Atome}},\ }\href {https://doi.org/10.1007/BF02980624}
  {\bibfield  {journal} {\bibinfo  {journal} {Z. Phys.}\ }\textbf {\bibinfo
  {volume} {31}},\ \bibinfo {pages} {681} (\bibinfo {year} {1925})}\BibitemShut
  {NoStop}%
\bibitem [{\citenamefont {Crowley}\ and\ \citenamefont
  {Gregori}(2014)}]{CROWLEY201455}%
  \BibitemOpen
  \bibfield  {author} {\bibinfo {author} {\bibfnamefont {B.}~\bibnamefont
  {Crowley}}\ and\ \bibinfo {author} {\bibfnamefont {G.}~\bibnamefont
  {Gregori}},\ }\bibfield  {title} {\bibinfo {title} {Quantum theory of thomson
  scattering},\ }\href
  {https://doi.org/https://doi.org/10.1016/j.hedp.2014.08.002} {\bibfield
  {journal} {\bibinfo  {journal} {High Energy Density Physics}\ }\textbf
  {\bibinfo {volume} {13}},\ \bibinfo {pages} {55} (\bibinfo {year}
  {2014})}\BibitemShut {NoStop}%
\bibitem [{\citenamefont {Wang}\ and\ \citenamefont
  {Zhu}(2020)}]{doi:10.1063/5.0011416}%
  \BibitemOpen
  \bibfield  {author} {\bibinfo {author} {\bibfnamefont {S.-X.}\ \bibnamefont
  {Wang}}\ and\ \bibinfo {author} {\bibfnamefont {L.-F.}\ \bibnamefont {Zhu}},\
  }\bibfield  {title} {\bibinfo {title} {Non-resonant inelastic x-ray
  scattering spectroscopy: A momentum probe to detect the electronic structures
  of atoms and molecules},\ }\href {https://doi.org/10.1063/5.0011416}
  {\bibfield  {journal} {\bibinfo  {journal} {Matter and Radiation at
  Extremes}\ }\textbf {\bibinfo {volume} {5}},\ \bibinfo {pages} {054201}
  (\bibinfo {year} {2020})},\ \Eprint
  {https://arxiv.org/abs/https://doi.org/10.1063/5.0011416}
  {https://doi.org/10.1063/5.0011416} \BibitemShut {NoStop}%
\bibitem [{\citenamefont {Price}\ and\ \citenamefont
  {Skold}(1986)}]{PRICE19861}%
  \BibitemOpen
  \bibfield  {author} {\bibinfo {author} {\bibfnamefont {D.~L.}\ \bibnamefont
  {Price}}\ and\ \bibinfo {author} {\bibfnamefont {K.}~\bibnamefont {Skold}},\
  }\bibfield  {title} {\bibinfo {title} {1. introduction to neutron
  scattering**this work supported by the u. s. department of energy},\ }in\
  \href {https://doi.org/https://doi.org/10.1016/S0076-695X(08)60554-2} {\emph
  {\bibinfo {booktitle} {Neutron Scattering}}},\ \bibinfo {series} {Methods in
  Experimental Physics}, Vol.~\bibinfo {volume} {23},\ \bibinfo {editor}
  {edited by\ \bibinfo {editor} {\bibfnamefont {K.}~\bibnamefont {Sköld}}\
  and\ \bibinfo {editor} {\bibfnamefont {D.~L.}\ \bibnamefont {Price}}}\
  (\bibinfo  {publisher} {Academic Press},\ \bibinfo {year} {1986})\ pp.\
  \bibinfo {pages} {1--97}\BibitemShut {NoStop}%
\bibitem [{\citenamefont {Nelin}\ and\ \citenamefont
  {Nilsson}(1972)}]{PhysRevB.5.3151}%
  \BibitemOpen
  \bibfield  {author} {\bibinfo {author} {\bibfnamefont {G.}~\bibnamefont
  {Nelin}}\ and\ \bibinfo {author} {\bibfnamefont {G.}~\bibnamefont
  {Nilsson}},\ }\bibfield  {title} {\bibinfo {title} {Phonon density of states
  in germanium at 80 k measured by neutron spectrometry},\ }\href
  {https://doi.org/10.1103/PhysRevB.5.3151} {\bibfield  {journal} {\bibinfo
  {journal} {Phys. Rev. B}\ }\textbf {\bibinfo {volume} {5}},\ \bibinfo {pages}
  {3151} (\bibinfo {year} {1972})}\BibitemShut {NoStop}%
\bibitem [{\citenamefont {Kamitakahara}\ \emph {et~al.}(1984)\citenamefont
  {Kamitakahara}, \citenamefont {Shanks}, \citenamefont {McClelland},
  \citenamefont {Buchenau}, \citenamefont {Gompf},\ and\ \citenamefont
  {Pintschovius}}]{PhysRevLett.52.644}%
  \BibitemOpen
  \bibfield  {author} {\bibinfo {author} {\bibfnamefont {W.~A.}\ \bibnamefont
  {Kamitakahara}}, \bibinfo {author} {\bibfnamefont {H.~R.}\ \bibnamefont
  {Shanks}}, \bibinfo {author} {\bibfnamefont {J.~F.}\ \bibnamefont
  {McClelland}}, \bibinfo {author} {\bibfnamefont {U.}~\bibnamefont
  {Buchenau}}, \bibinfo {author} {\bibfnamefont {F.}~\bibnamefont {Gompf}},\
  and\ \bibinfo {author} {\bibfnamefont {L.}~\bibnamefont {Pintschovius}},\
  }\bibfield  {title} {\bibinfo {title} {Measurement of phonon densities of
  states for pure and hydrogenated amorphous silicon},\ }\href
  {https://doi.org/10.1103/PhysRevLett.52.644} {\bibfield  {journal} {\bibinfo
  {journal} {Phys. Rev. Lett.}\ }\textbf {\bibinfo {volume} {52}},\ \bibinfo
  {pages} {644} (\bibinfo {year} {1984})}\BibitemShut {NoStop}%
\bibitem [{\citenamefont {Blakemore}(1982)}]{doi:10.1063/1.331665}%
  \BibitemOpen
  \bibfield  {author} {\bibinfo {author} {\bibfnamefont {J.~S.}\ \bibnamefont
  {Blakemore}},\ }\bibfield  {title} {\bibinfo {title} {Semiconducting and
  other major properties of gallium arsenide},\ }\href
  {https://doi.org/10.1063/1.331665} {\bibfield  {journal} {\bibinfo  {journal}
  {Journal of Applied Physics}\ }\textbf {\bibinfo {volume} {53}},\ \bibinfo
  {pages} {R123} (\bibinfo {year} {1982})},\ \Eprint
  {https://arxiv.org/abs/https://doi.org/10.1063/1.331665}
  {https://doi.org/10.1063/1.331665} \BibitemShut {NoStop}%
\bibitem [{\citenamefont {BOUHADDA}\ \emph {et~al.}(2012)\citenamefont
  {BOUHADDA}, \citenamefont {BENTABET}, \citenamefont {FENINECHE},\ and\
  \citenamefont {BOUDOUMA}}]{doi:10.1142/S2047684112500261}%
  \BibitemOpen
  \bibfield  {author} {\bibinfo {author} {\bibfnamefont {Y.}~\bibnamefont
  {BOUHADDA}}, \bibinfo {author} {\bibfnamefont {A.}~\bibnamefont {BENTABET}},
  \bibinfo {author} {\bibfnamefont {N.~E.}\ \bibnamefont {FENINECHE}},\ and\
  \bibinfo {author} {\bibfnamefont {Y.}~\bibnamefont {BOUDOUMA}},\ }\bibfield
  {title} {\bibinfo {title} {The ab initio calculation of the dynamical and the
  thermodynamic properties of the zinc-blende gax (x=n, p, as and sb)},\ }\href
  {https://doi.org/10.1142/S2047684112500261} {\bibfield  {journal} {\bibinfo
  {journal} {International Journal of Computational Materials Science and
  Engineering}\ }\textbf {\bibinfo {volume} {01}},\ \bibinfo {pages} {1250026}
  (\bibinfo {year} {2012})},\ \Eprint
  {https://arxiv.org/abs/https://doi.org/10.1142/S2047684112500261}
  {https://doi.org/10.1142/S2047684112500261} \BibitemShut {NoStop}%
\bibitem [{\citenamefont {Kaur}\ and\ \citenamefont
  {Sinha}(2020)}]{doi:10.1063/5.0017269}%
  \BibitemOpen
  \bibfield  {author} {\bibinfo {author} {\bibfnamefont {T.}~\bibnamefont
  {Kaur}}\ and\ \bibinfo {author} {\bibfnamefont {M.~M.}\ \bibnamefont
  {Sinha}},\ }\bibfield  {title} {\bibinfo {title} {First principle study of
  structural, electronic and vibrational properties of 3c-sic},\ }\href
  {https://doi.org/10.1063/5.0017269} {\bibfield  {journal} {\bibinfo
  {journal} {AIP Conference Proceedings}\ }\textbf {\bibinfo {volume} {2265}},\
  \bibinfo {pages} {030384} (\bibinfo {year} {2020})},\ \Eprint
  {https://arxiv.org/abs/https://aip.scitation.org/doi/pdf/10.1063/5.0017269}
  {https://aip.scitation.org/doi/pdf/10.1063/5.0017269} \BibitemShut {NoStop}%
\bibitem [{\citenamefont {Petretto}\ \emph {et~al.}(2018)\citenamefont
  {Petretto}, \citenamefont {Dwaraknath}, \citenamefont {P.C.~Miranda},
  \citenamefont {Winston}, \citenamefont {Giantomassi}, \citenamefont {van
  Setten}, \citenamefont {Gonze}, \citenamefont {Persson}, \citenamefont
  {Hautier},\ and\ \citenamefont {Rignanese}}]{Petretto2018}%
  \BibitemOpen
  \bibfield  {author} {\bibinfo {author} {\bibfnamefont {G.}~\bibnamefont
  {Petretto}}, \bibinfo {author} {\bibfnamefont {S.}~\bibnamefont
  {Dwaraknath}}, \bibinfo {author} {\bibfnamefont {H.}~\bibnamefont
  {P.C.~Miranda}}, \bibinfo {author} {\bibfnamefont {D.}~\bibnamefont
  {Winston}}, \bibinfo {author} {\bibfnamefont {M.}~\bibnamefont
  {Giantomassi}}, \bibinfo {author} {\bibfnamefont {M.~J.}\ \bibnamefont {van
  Setten}}, \bibinfo {author} {\bibfnamefont {X.}~\bibnamefont {Gonze}},
  \bibinfo {author} {\bibfnamefont {K.~A.}\ \bibnamefont {Persson}}, \bibinfo
  {author} {\bibfnamefont {G.}~\bibnamefont {Hautier}},\ and\ \bibinfo {author}
  {\bibfnamefont {G.-M.}\ \bibnamefont {Rignanese}},\ }\bibfield  {title}
  {\bibinfo {title} {High-throughput density-functional perturbation theory
  phonons for inorganic materials},\ }\href
  {https://doi.org/10.1038/sdata.2018.65} {\bibfield  {journal} {\bibinfo
  {journal} {Scientific Data}\ }\textbf {\bibinfo {volume} {5}},\ \bibinfo
  {pages} {180065} (\bibinfo {year} {2018})}\BibitemShut {NoStop}%
\bibitem [{\citenamefont {Serrano}\ \emph {et~al.}(2002)\citenamefont
  {Serrano}, \citenamefont {Strempfer}, \citenamefont {Cardona}, \citenamefont
  {Schwoerer-Böhning}, \citenamefont {Requardt}, \citenamefont {Lorenzen},
  \citenamefont {Stojetz}, \citenamefont {Pavone},\ and\ \citenamefont
  {Choyke}}]{doi:10.1063/1.1484241}%
  \BibitemOpen
  \bibfield  {author} {\bibinfo {author} {\bibfnamefont {J.}~\bibnamefont
  {Serrano}}, \bibinfo {author} {\bibfnamefont {J.}~\bibnamefont {Strempfer}},
  \bibinfo {author} {\bibfnamefont {M.}~\bibnamefont {Cardona}}, \bibinfo
  {author} {\bibfnamefont {M.}~\bibnamefont {Schwoerer-Böhning}}, \bibinfo
  {author} {\bibfnamefont {H.}~\bibnamefont {Requardt}}, \bibinfo {author}
  {\bibfnamefont {M.}~\bibnamefont {Lorenzen}}, \bibinfo {author}
  {\bibfnamefont {B.}~\bibnamefont {Stojetz}}, \bibinfo {author} {\bibfnamefont
  {P.}~\bibnamefont {Pavone}},\ and\ \bibinfo {author} {\bibfnamefont {W.~J.}\
  \bibnamefont {Choyke}},\ }\bibfield  {title} {\bibinfo {title} {Determination
  of the phonon dispersion of zinc blende (3c) silicon carbide by inelastic
  x-ray scattering},\ }\href {https://doi.org/10.1063/1.1484241} {\bibfield
  {journal} {\bibinfo  {journal} {Applied Physics Letters}\ }\textbf {\bibinfo
  {volume} {80}},\ \bibinfo {pages} {4360} (\bibinfo {year} {2002})},\ \Eprint
  {https://arxiv.org/abs/https://doi.org/10.1063/1.1484241}
  {https://doi.org/10.1063/1.1484241} \BibitemShut {NoStop}%
\bibitem [{\citenamefont {Armengaud}\ \emph {et~al.}(2018)\citenamefont
  {Armengaud} \emph {et~al.}}]{EDELWEISS:2018tde}%
  \BibitemOpen
  \bibfield  {author} {\bibinfo {author} {\bibfnamefont {E.}~\bibnamefont
  {Armengaud}} \emph {et~al.} (\bibinfo {collaboration} {EDELWEISS}),\
  }\bibfield  {title} {\bibinfo {title} {{Searches for electron interactions
  induced by new physics in the EDELWEISS-III Germanium bolometers}},\ }\href
  {https://doi.org/10.1103/PhysRevD.98.082004} {\bibfield  {journal} {\bibinfo
  {journal} {Phys. Rev. D}\ }\textbf {\bibinfo {volume} {98}},\ \bibinfo
  {pages} {082004} (\bibinfo {year} {2018})},\ \Eprint
  {https://arxiv.org/abs/1808.02340} {arXiv:1808.02340 [hep-ex]} \BibitemShut
  {NoStop}%
\bibitem [{\citenamefont {Agnese}\ \emph {et~al.}(2017)\citenamefont {Agnese}
  \emph {et~al.}}]{SuperCDMS:2016wui}%
  \BibitemOpen
  \bibfield  {author} {\bibinfo {author} {\bibfnamefont {R.}~\bibnamefont
  {Agnese}} \emph {et~al.} (\bibinfo {collaboration} {SuperCDMS}),\ }\bibfield
  {title} {\bibinfo {title} {{Projected Sensitivity of the SuperCDMS SNOLAB
  experiment}},\ }\href {https://doi.org/10.1103/PhysRevD.95.082002} {\bibfield
   {journal} {\bibinfo  {journal} {Phys. Rev. D}\ }\textbf {\bibinfo {volume}
  {95}},\ \bibinfo {pages} {082002} (\bibinfo {year} {2017})},\ \Eprint
  {https://arxiv.org/abs/1610.00006} {arXiv:1610.00006 [physics.ins-det]}
  \BibitemShut {NoStop}%
\bibitem [{\citenamefont {Trickle}\ \emph
  {et~al.}(2020{\natexlab{b}})\citenamefont {Trickle}, \citenamefont {Zhang},
  \citenamefont {Zurek}, \citenamefont {Inzani},\ and\ \citenamefont
  {Griffin}}]{Trickle_2020}%
  \BibitemOpen
  \bibfield  {author} {\bibinfo {author} {\bibfnamefont {T.}~\bibnamefont
  {Trickle}}, \bibinfo {author} {\bibfnamefont {Z.}~\bibnamefont {Zhang}},
  \bibinfo {author} {\bibfnamefont {K.~M.}\ \bibnamefont {Zurek}}, \bibinfo
  {author} {\bibfnamefont {K.}~\bibnamefont {Inzani}},\ and\ \bibinfo {author}
  {\bibfnamefont {S.~M.}\ \bibnamefont {Griffin}},\ }\bibfield  {title}
  {\bibinfo {title} {Multi-channel direct detection of light dark matter:
  theoretical framework},\ }\bibfield  {journal} {\bibinfo  {journal} {Journal
  of High Energy Physics}\ }\textbf {\bibinfo {volume} {2020}},\ \href
  {https://doi.org/10.1007/jhep03(2020)036} {10.1007/jhep03(2020)036} (\bibinfo
  {year} {2020}{\natexlab{b}})\BibitemShut {NoStop}%
\bibitem [{\citenamefont {Knapen}\ \emph
  {et~al.}(2021{\natexlab{a}})\citenamefont {Knapen}, \citenamefont
  {Kozaczuk},\ and\ \citenamefont {Lin}}]{Knapen:2021bwg}%
  \BibitemOpen
  \bibfield  {author} {\bibinfo {author} {\bibfnamefont {S.}~\bibnamefont
  {Knapen}}, \bibinfo {author} {\bibfnamefont {J.}~\bibnamefont {Kozaczuk}},\
  and\ \bibinfo {author} {\bibfnamefont {T.}~\bibnamefont {Lin}},\ }\bibfield
  {title} {\bibinfo {title} {{DarkELF: A python package for dark matter
  scattering in dielectric targets}},\ }\href@noop {} {\  (\bibinfo {year}
  {2021}{\natexlab{a}})},\ \Eprint {https://arxiv.org/abs/2104.12786}
  {arXiv:2104.12786 [hep-ph]} \BibitemShut {NoStop}%
\bibitem [{\citenamefont {Lewin}\ and\ \citenamefont
  {Smith}(1996)}]{LEWIN199687}%
  \BibitemOpen
  \bibfield  {author} {\bibinfo {author} {\bibfnamefont {J.}~\bibnamefont
  {Lewin}}\ and\ \bibinfo {author} {\bibfnamefont {P.}~\bibnamefont {Smith}},\
  }\bibfield  {title} {\bibinfo {title} {Review of mathematics, numerical
  factors, and corrections for dark matter experiments based on elastic nuclear
  recoil},\ }\href
  {https://doi.org/https://doi.org/10.1016/S0927-6505(96)00047-3} {\bibfield
  {journal} {\bibinfo  {journal} {Astroparticle Physics}\ }\textbf {\bibinfo
  {volume} {6}},\ \bibinfo {pages} {87} (\bibinfo {year} {1996})}\BibitemShut
  {NoStop}%
\bibitem [{\citenamefont {Hochberg}\ \emph
  {et~al.}(2021{\natexlab{b}})\citenamefont {Hochberg}, \citenamefont {Kahn},
  \citenamefont {Kurinsky}, \citenamefont {Lehmann}, \citenamefont {Yu},\ and\
  \citenamefont {Berggren}}]{Hochberg:2021pkt}%
  \BibitemOpen
  \bibfield  {author} {\bibinfo {author} {\bibfnamefont {Y.}~\bibnamefont
  {Hochberg}}, \bibinfo {author} {\bibfnamefont {Y.}~\bibnamefont {Kahn}},
  \bibinfo {author} {\bibfnamefont {N.}~\bibnamefont {Kurinsky}}, \bibinfo
  {author} {\bibfnamefont {B.~V.}\ \bibnamefont {Lehmann}}, \bibinfo {author}
  {\bibfnamefont {T.~C.}\ \bibnamefont {Yu}},\ and\ \bibinfo {author}
  {\bibfnamefont {K.~K.}\ \bibnamefont {Berggren}},\ }\bibfield  {title}
  {\bibinfo {title} {{Determining Dark Matter-Electron Scattering Rates from
  the Dielectric Function}},\ }\href@noop {} {\  (\bibinfo {year}
  {2021}{\natexlab{b}})},\ \Eprint {https://arxiv.org/abs/2101.08263}
  {arXiv:2101.08263 [hep-ph]} \BibitemShut {NoStop}%
\bibitem [{\citenamefont {Knapen}\ \emph
  {et~al.}(2021{\natexlab{b}})\citenamefont {Knapen}, \citenamefont
  {Kozaczuk},\ and\ \citenamefont {Lin}}]{Knapen:2021run}%
  \BibitemOpen
  \bibfield  {author} {\bibinfo {author} {\bibfnamefont {S.}~\bibnamefont
  {Knapen}}, \bibinfo {author} {\bibfnamefont {J.}~\bibnamefont {Kozaczuk}},\
  and\ \bibinfo {author} {\bibfnamefont {T.}~\bibnamefont {Lin}},\ }\bibfield
  {title} {\bibinfo {title} {{Dark matter-electron scattering in
  dielectrics}},\ }\href {https://doi.org/10.1103/PhysRevD.104.015031}
  {\bibfield  {journal} {\bibinfo  {journal} {Phys. Rev. D}\ }\textbf {\bibinfo
  {volume} {104}},\ \bibinfo {pages} {015031} (\bibinfo {year}
  {2021}{\natexlab{b}})},\ \Eprint {https://arxiv.org/abs/2101.08275}
  {arXiv:2101.08275 [hep-ph]} \BibitemShut {NoStop}%
\bibitem [{Ber(2021)}]{BerghausGithub}%
  \BibitemOpen
  \bibfield  {title} {\bibinfo {title}
  {https://github.com/kberghaus/phonon$\_$background},\ }\href@noop {} {\
  (\bibinfo {year} {2021})}\BibitemShut {NoStop}%
\bibitem [{\citenamefont {Schober}(2014)}]{schober_2014}%
  \BibitemOpen
  \bibfield  {author} {\bibinfo {author} {\bibfnamefont {H.}~\bibnamefont
  {Schober}},\ }\bibfield  {title} {\bibinfo {title} {An introduction to the
  theory of nuclear neutron scattering in condensed matter},\ }\href
  {https://doi.org/10.3233/jnr-140016} {\bibfield  {journal} {\bibinfo
  {journal} {Journal of Neutron Research}\ }\textbf {\bibinfo {volume} {17}},\
  \bibinfo {pages} {109–357} (\bibinfo {year} {2014})}\BibitemShut {NoStop}%
\bibitem [{\citenamefont {Scorza}(2015)}]{Scorza:2015vla}%
  \BibitemOpen
  \bibfield  {author} {\bibinfo {author} {\bibfnamefont {S.}~\bibnamefont
  {Scorza}} (\bibinfo {collaboration} {EDELWEISS}),\ }\bibfield  {title}
  {\bibinfo {title} {{Background investigation in EDELWEISS-III}},\ }\href
  {https://doi.org/10.1063/1.4928002} {\bibfield  {journal} {\bibinfo
  {journal} {AIP Conf. Proc.}\ }\textbf {\bibinfo {volume} {1672}},\ \bibinfo
  {pages} {100002} (\bibinfo {year} {2015})}\BibitemShut {NoStop}%
\bibitem [{\citenamefont {Essig}\ \emph {et~al.}(2018)\citenamefont {Essig},
  \citenamefont {Sholapurkar},\ and\ \citenamefont {Yu}}]{Essig:2018tss}%
  \BibitemOpen
  \bibfield  {author} {\bibinfo {author} {\bibfnamefont {R.}~\bibnamefont
  {Essig}}, \bibinfo {author} {\bibfnamefont {M.}~\bibnamefont {Sholapurkar}},\
  and\ \bibinfo {author} {\bibfnamefont {T.-T.}\ \bibnamefont {Yu}},\
  }\bibfield  {title} {\bibinfo {title} {{Solar Neutrinos as a Signal and
  Background in Direct-Detection Experiments Searching for Sub-GeV Dark Matter
  With Electron Recoils}},\ }\href {https://doi.org/10.1103/PhysRevD.97.095029}
  {\bibfield  {journal} {\bibinfo  {journal} {Phys. Rev. D}\ }\textbf {\bibinfo
  {volume} {97}},\ \bibinfo {pages} {095029} (\bibinfo {year} {2018})},\
  \Eprint {https://arxiv.org/abs/1801.10159} {arXiv:1801.10159 [hep-ph]}
  \BibitemShut {NoStop}%
\bibitem [{\citenamefont {Bahcall}\ \emph {et~al.}(2005)\citenamefont
  {Bahcall}, \citenamefont {Serenelli},\ and\ \citenamefont
  {Basu}}]{Bahcall:2004pz}%
  \BibitemOpen
  \bibfield  {author} {\bibinfo {author} {\bibfnamefont {J.~N.}\ \bibnamefont
  {Bahcall}}, \bibinfo {author} {\bibfnamefont {A.~M.}\ \bibnamefont
  {Serenelli}},\ and\ \bibinfo {author} {\bibfnamefont {S.}~\bibnamefont
  {Basu}},\ }\bibfield  {title} {\bibinfo {title} {{New solar opacities,
  abundances, helioseismology, and neutrino fluxes}},\ }\href
  {https://doi.org/10.1086/428929} {\bibfield  {journal} {\bibinfo  {journal}
  {Astrophys. J. Lett.}\ }\textbf {\bibinfo {volume} {621}},\ \bibinfo {pages}
  {L85} (\bibinfo {year} {2005})},\ \Eprint
  {https://arxiv.org/abs/astro-ph/0412440} {arXiv:astro-ph/0412440}
  \BibitemShut {NoStop}%
\bibitem [{\citenamefont {{Lewin}}\ and\ \citenamefont
  {{Smith}}(1996)}]{LewinSmith}%
  \BibitemOpen
  \bibfield  {author} {\bibinfo {author} {\bibfnamefont {J.~D.}\ \bibnamefont
  {{Lewin}}}\ and\ \bibinfo {author} {\bibfnamefont {P.~F.}\ \bibnamefont
  {{Smith}}},\ }\bibfield  {title} {\bibinfo {title} {{Review of mathematics,
  numerical factors, and corrections for dark matter experiments based on
  elastic nuclear recoil}},\ }\href
  {https://doi.org/10.1016/S0927-6505(96)00047-3} {\bibfield  {journal}
  {\bibinfo  {journal} {Astroparticle Physics}\ }\textbf {\bibinfo {volume}
  {6}},\ \bibinfo {pages} {87} (\bibinfo {year} {1996})}\BibitemShut {NoStop}%
\end{thebibliography}%

\end{document}